\documentclass[manuscript]{aastex}

\usepackage{bm}
\usepackage{color}
\usepackage{ulem}
\usepackage{graphicx}

\shorttitle{Oscillations excited by plasmoids formed during magnetic
reconnection} \shortauthors{Jel\'\i nek et al.}

\begin{document}


\title{Oscillations excited by plasmoids formed during magnetic reconnection in a vertical gravitationally stratified current sheet}


\author{P. Jel\'\i nek\altaffilmark{1}}
\affil{University of South Bohemia, Faculty of Science, Institute of Physics and Biophysics, Brani\v sovsk\'a 10, CZ -- 370 05 \v{C}esk\'e
              Bud\v{e}jovice, Czech Republic}
\email{pjelinek@prf.jcu.cz}

\author{M. Karlick\'y\altaffilmark{2}}
\affil{Czech Academy of Sciences, v.~v.~i., Astronomical Institute, Fri\v cova 258, CZ -- 251 65 Ond\v rejov, Czech
             Republic}

\author{T. Van Doorsselaere\altaffilmark{3}}
\affil{Centre for mathematical Plasma Astrophysics (CmPA), Mathematics Department, KU Leuven, Celestijnenlaan 200B bus 2400, 3001 Leuven, Belgium}

\author{M. B\'arta\altaffilmark{2}}
\affil{Czech Academy of Sciences, v.~v.~i., Astronomical Institute, Fri\v cova 258, CZ -- 251 65 Ond\v rejov, Czech
             Republic}


\begin{abstract}
Using the FLASH code, which solves the full set of the two-dimensional (2-D)
non-ideal (resistive) time-dependent magnetohydrodynamic (MHD) equations, we
study processes during the magnetic reconnection in a vertical gravitationally
stratified current sheet. We show that during these processes, which correspond
to processes in solar flares, plasmoids are formed due to the tearing mode
instability of the current sheet. These plasmoids move upwards or downwards
along the vertical current sheet, and some of them merge into larger plasmoids.
We study the density and temperature structure of these plasmoids and their
time evolution in details. We found that during the merging of two plasmoids
the resulting larger plasmoid starts to oscillate; in our model with a $\sim
25~\mathrm{s}$ period. On the other hand, the plasmoid moving downwards merges
with the underlying flare arcade which also starts to oscillate during this
process; in our model with a $\sim 35~\mathrm{s}$ period. It is shown that the
merging process of plasmoid with the flare arcade is a complex process as
presented by complex density and temperature structures of the oscillating
arcade. Moreover, all these processes are associated with magnetoacoustic waves
produced by the  motion and merging of plasmoids.
\end{abstract}



\section{Introduction}
Magnetohydrodynamic (MHD) plasma waves and oscillations are omnipresent in the solar atmosphere and play a very important role in many phenomena observed in solar and stellar atmospheres \citep{2004psci.book.....A,2012scsd.book.....S} as well as in the problem of the acceleration of the fast component of the solar wind \citep[e.g.][]{2004JGRA..109.7103L,2008ApJ...682..667L,2016SSRv..200...75N}.

The magnetically dominated solar plasma supports the propagation of various types of MHD waves \citep[e.g.][]{2005LRSP....2....3N,2012RSPTA.370.3193D,2016ApJ...833...51Y}. In solar flares, quasi-periodic pulsations (QPPs) are commonly observed in radio, soft X-ray, hard X-ray, ultraviolet, and even in gamma-ray emissions \citep{1984ApJ...279..857R,2003SoPh..218..183F,2005A&A...435..753W,2006A&A...452..343N,2010PPCF...52l4009N,2016SSRv..200...75N}.

The period of these oscillations ranges from sub-seconds to tens of minutes \citep{2006A&A...460..865M,2008SoPh..253..117T,2010SoPh..261..281K,2010SoPh..267..329K,2014ApJ...791...44H,2014A&A...569A..12N}, and some of them were interpreted as manifestations of the transverse, kink mode \citep[e.g.][]{2013SoPh..284..559K} or sausage modes \citep[e.g.][]{2003A&A...412L...7N}, collectively supported by flare loops. When interpreted this way, the measured QPPs can be seismologically exploited to infer such key parameters as the magnetic field strength in the key region where flare energy is released \citep[e.g.][]{2015ApJ...812...22C,  2016SoPh..291..877G, 2016ApJ...833..114C}. Several theoretical models have been proposed to explain the generation of these QPPs~\citep{2009SSRv..149..119N, 2016SoPh..291.3143V}. Quasi-periodic pulsations are often detected in almost all wavelenghts -- microwave, white light, X-ray, and gamma-ray bands. The periods of QPPs are sub-second to tens of seconds. The damping time of QPPs is not yet very well measured but is around a few periods \citep[e.g.][]{2015A&A...574A..53K}.

Despite many theoretical models and the numerical results, it is not well understood how QPPS are generated and the mechanisms are still debated in the literature. For the review of recent numerical results see e.g. \citet{2014RAA....14..805P}. Here we study in detail the generation of transverse waves in a magnetic arcade impacted by plasmoids ejected from a current sheet. It deviates from the work by \citet{2015ApJ...805..135T} and \citet{2016ApJ...823..150T} in the sense that they consider a steady reconnection regime which leads to (dynamically evolving) shocks in the post-flare arcade, whereas we consider a bursty outflow from the reconnection site (in the form of plasmoids), which then excites transverse waves. \citet{2016ApJ...823..150T}  imposed a localized resistivity in their simulation model. This localized resistivity is fixed in time and space to realize a fast and quasi-steady magnetic  reconnection with a single X-point. This means that they neglected the oscillations caused by plasmoids and focused on the oscillations excited by a quasi-steady reconnection outflow. Our simulation is thus complementary to their simulation because we focus just on oscillations generated during  mutual interactions of plasmoids and interaction of plasmoids with the flare arcade. An excitation of the fast magnetoacoustic waves by ejected plasmoids after their interaction with ambient magnetic field was also investigated by \cite{2015ApJ...800..111Y}. However, in contrast with our present work, they performed their calculations in the so-called interchange-reconnection scenario.

Apart from the fact that the transverse waves excited by plasmoids may be the mechanism behind the observed QPPs in solar flares, they could also be important for the energy transport in solar flares. It was pointed out by \citet{2008ApJ...675.1645F} that magnetic tension waves can play an important role in the transport of energy from the flare site to the lower atmospheric layers. The energy of the transverse waves is then transported down and dissipated in the chromosphere \citep{2013ApJ...765...81R,2016ApJ...818L..20R}, where it can be used to accelerate energetic particles.

In our previous papers \citep{2012A&A...537A..46J,2012A&A...546A..49J,2015A&A...581A.131J} we numerically investigated the behaviour of fast magnetoacoustic waves in a Harris current sheet structure but we did not consider resistivity. In the present paper we included the resistivity to show not only the formation of plasmoids and their coalescence, similarly as in the paper by \cite{2008A&A...477..649B}, but also the collision of plasmoids outflowing from the reconnection site with the flare magnetic arcade, triggering oscillatory processes. It was shown previously by \citet{2015ApJ...810L..19I} that Particle-In-Cell (PIC) simulations result in the formation of several plasmoids, until a large plasmoid stabilises the system.

In real flares, for a sufficiently fast reconnection process, anomalous
resistivity needs to be considered. This resistivity is generated at locations
where the drift of electrons, forming the electric current, overcomes some
critical velocity (e.g., the ion-sound or even electron thermal velocity)
\citep{1978A&A....68..145N}. In solar flare conditions it requires a very
narrow current sheet $(10 - 100~\mathrm{m})$. Thus, in any global MHD model of
solar flares it is difficult to consider such a very narrow current sheet and
simultaneously high values of the very localized anomalous resistivity. Due to
this problem, in the present paper we made a compromise, we take the current
sheet that is much broader than that in a real flare and the resistivity which
is greater than the collisional resistivity, but lower than the anomalous one.
In order to be in agreement with observed oscillations of the flare
arcade and plasmoid~\citep{2006A&A...446..675V,2010SoPh..266...71K}, we have
chosen the resistivity 100 times greater than the collisional one. Although in
several aspects the model can be improved, it presents the processes that are
new and important, and still not considered in connection with solar flares.

\cite{1987ApJ...321.1031T} studied the merger process of two plasmoids and they found that during this process a larger plasmoid is formed and this plasmoid oscillates with the period of about the Alfv\'en transit time, i.e., $P \approx L/c_\mathrm{A}$, where $L$ is the characteristic width of the plasmoid and $c_\mathrm{A}$ is the mean Alfv\'en speed in the plasmoid. Plasmoids in solar flares have been observed in soft X-rays and EUV \citep{1998ApJ...499..934O,2007A&A...475..685K,2012ApJ...745L...6T,2012ApJ...754...13S}. In some cases processes connected with these plasmoids produce the so called drifting pulsating structures (DPSs) in the decimetric range of radio waves \citep{2000A&A...360..715K,2008A&A...477..649B,2011ApJ...737...24B,2011ApJ...733..107K}.

The paper is structured as follows: In Section 2 we present our numerical model, including the initial equilibrium and perturbations. The results of the numerical simulations and their interpretation are summarized in Section 3. Finally, we complete the paper by some conclusions in the last Sect. 4.

\section{Model}
\subsection{Governing equations}

Our numerical model describes a gravitationally stratified solar atmosphere, in which the plasma dynamics are described by the 2-D, time-dependent non-ideal (resistive) MHD equations. We use the FLASH code~\citep{2009JCoPh.228..952L,2013JCoPh.243..269L}, where the MHD equations are formulated in conservative form as
\begin{equation}\label{eq1}
\frac{\partial \varrho}{\partial t} + \nabla \cdot(\varrho \bm{\mathrm{v}}) = 0,
\end{equation}
\begin{equation}\label{eq2}
\frac{\partial \varrho \bm{\mathrm{v}}}{\partial t} + \nabla \cdot \left(\varrho \bm{\mathrm{v}} \bm{\mathrm{v}} -\bm{\mathrm{B}}\bm{\mathrm{B}}\right) + \nabla p_{*}=\varrho \bm{\mathrm{g}},
\end{equation}
\begin{eqnarray}\label{eq3}
\frac{\partial \varrho E}{\partial t} + \nabla \cdot \left[\left(\varrho E+p_{*}\right)\bm{\mathrm{v}} - \bm{\mathrm{B}}(\bm{\mathrm{v}} \cdot \bm{\mathrm{B}})\right] =  \nonumber \\
= \varrho \bm{\mathrm{g}} \cdot \bm{\mathrm{v}} + \nabla \cdot (\bm{\mathrm{B}} \times (\eta \nabla \times \bm{\mathrm{B}})),
\end{eqnarray}
\begin{equation}\label{eq4}
\frac{\partial \bm{\mathrm{B}}}{\partial t} + \nabla \cdot (\bm{\mathrm{v}}\bm{\mathrm{B}}-\bm{\mathrm{B}}\bm{\mathrm{v}})=-\nabla \times (\eta \nabla \times \bm{\mathrm{B}}),
\end{equation}
\begin{equation}\label{eq5}
\nabla\cdot\bm{\mathrm{B}}=0.
\end{equation}
Here $\varrho$ is the mass density, $\bm{\mathrm{v}}$ is the flow velocity, $\bm{\mathrm{B}}$ is the magnetic field strength, $\bm{\mathrm{g}} = [0,-g_{\sun},0]$ is the gravitational acceleration with $g_{\sun} = 274~\mathrm{m s^{-2}}$ and $\eta$ is the magnetic diffusivity. According to \cite{1982soma.book.....P}, the magnetic diffusivity in the collisional regime can be expressed as $\eta = 10^{9}T^{-3/2}$, which gives us for the temperature $T\approx 10^6~\mathrm{K}$ a value $1~\mathrm{m^{2} \cdot s^{-1}}$. To see the magnetic reconnection in our models at reasonable times, we tested different values of the magnetic diffusivity and finally used the value which is $100$ times higher than the collisional diffusivity, but much smaller than the anomalous one. The total pressure $p_{*}$ is given by:
\begin{equation}\label{eq6}
p_{*} = \left(p + \frac{B^2}{2 \mu_0}\right),
\end{equation}
$p$ is the fluid thermal pressure, $B$ is the magnitude of the magnetic field.
The specific total energy $E$ in Eq. (\ref{eq3}) is expressed as:
\begin{equation}\label{eq7}
E = \epsilon + \frac{v^2}{2} + \frac{B^2}{2 \mu_0 \varrho},
\end{equation}
where $\epsilon$ is the specific internal energy:
\begin{equation}\label{eq8}
\epsilon = \frac{p}{(\gamma-1)\varrho},
\end{equation}
with the adiabatic coefficient $\gamma = 5/3$, $v$ is the magnitude of the flow velocity and $\mu_0 = 1.26 \times 10^{-6}~\mathrm{H m}^{-1}$ is the magnetic permeability of free space.

Generally, the terms expressing the radiative losses $R_{\rm{loss}}$, thermal conduction $T_{\rm{cond}}$ and heating $H$ should be added to the set of MHD equations. In the initial state it is assumed that the radiative losses and thermal conduction are fully compensated by the heating $H$, i.e., $R_{\rm{loss}} + T_{\rm{cond}} + H = 0$, otherwise the unperturbed atmosphere is not in equilibrium, e.g., due to the steep temperature gradient in the transition region. Problems appear when the atmosphere is perturbed. Namely, there is no simple expression for the heating term $H$, which in the unperturbed atmosphere fully compensates $R_{\rm{loss}}$ and $T_{\rm{cond}}$, and in the perturbed atmosphere correctly describes the heating. Therefore, for the purpose of our study, we assume that $R_{\rm{loss}} + T_{\rm{cond}} + H = 0$ is valid during the whole studied processes.

\subsection{Initial equilibrium}
For a static ($\bm{v} = \bm{0}$) equilibrium, the Lorentz and gravity forces must be balanced by the pressure gradient in the entire physical domain
\begin{equation}\label{eq9}
-\nabla p+\bm{\mathrm{j}}\times\bm{\mathrm{B}} + \varrho \bm{\mathrm{g}} = \bm{0}.
\end{equation}
The solenoidal condition, $\nabla\cdot\bm{\mathrm{B}}=0$, is identically satisfied by using a
magnetic flux function, $\bm{\mathrm{A}}$,
\begin{equation}\label{eq10}
\bm{\mathrm{B}} = \nabla \times \bm{\mathrm{A}}.
\end{equation}
For calculating the magnetic field in the vertically oriented current sheet, we use the magnetic flux function $\bm{\mathrm{A}} = [0,0,A_z]$ as e.g. \cite{2012A&A...546A..49J,2015A&A...581A.131J}
\begin{equation}\label{eq11}
A_z = -B_{\mathrm{0}} w_\mathrm{cs} \ln\left\{\left|\cosh\left(\frac{x}{w_\mathrm{cs}}\right)\right|\right\}\exp \left(-\frac{y}{\lambda}\right).\label{eq:potential}
\end{equation}
Here the coefficient $\lambda \approx 100~\mathrm{Mm}$ denotes the magnetic scale-height. The symbol $B_{\mathrm{0}}$ is used for the external magnetic field and $w_\mathrm{cs}$ is the half width of the current-sheet. We set $w_\mathrm{cs} = 0.15~\mathrm{Mm}$. Note, that the prescription for $\bm{\mathrm{g}}$ and $\bm{\mathrm{A}}$ implies a current-sheet in the vertical direction $y$ with the current density flowing in the invariant direction $z$; $x$ being the horizontal direction perpendicular to the current-sheet.

The equilibrium gas pressure and mass density are computed according to the following equations, see e.g. \cite{2010ARep...54...86S} and \cite{2015ApJ...812..105J}:
\begin{equation}\label{eq12}
p(x,y) = p_{\rm h}(y) - \frac{1}{\mu_0}\left[\int\limits_{-\infty}^{x}\frac{\partial^2 A}{\partial y^2}\frac{\partial A}{\partial x}\mathrm{d}x +
\frac{1}{2}\left(\frac{\partial A}{\partial x}\right)^2\right],
\end{equation}

\begin{eqnarray}\label{eq13}
\varrho(x,y) &=& \varrho_{\rm h}(y) + \frac{1}{\mu_0 g_{\sun}}\Bigg\{\frac{\partial}{\partial{y}}\Bigg[\int\limits_{-\infty}^{x} \frac{\partial^2 A}{\partial y^2}\frac{\partial A}{\partial x}\mathrm{d}x + \nonumber \\
&+& \frac{1}{2}\Bigg(\frac{\partial A}{\partial x}\Bigg)^2\Bigg] - \frac{\partial A}{\partial y} \nabla^2 A\Bigg\}.
\end{eqnarray}
With the use of Eq.~(\ref{eq11}) in these general formulas we obtain the expressions for the equilibrium gas pressure
\begin{eqnarray}\label{eq14}
p(x,y) &=& p_h(y) - \frac{B_0^2}{2 \mu_0}\Bigg\{\frac{w_\mathrm{cs}^2}{\lambda^2}\ln^2\Bigg[\cosh\Bigg(\frac{x}{w_\mathrm{cs}}\Bigg)\Bigg] + \nonumber \\
&+& \tanh^2\left(\frac{x}{w_\mathrm{cs}}\right)\Bigg\}\exp\Bigg(-\frac{2y}{\lambda}\Bigg),
\end{eqnarray}
and mass density
\begin{eqnarray}\label{eq15}
&&\varrho(x,y) = \varrho_h(y) - \frac{B_0^2}{\mu_0 g_{\sun} \lambda}\Bigg\{\tanh^2\left(\frac{x}{w_\mathrm{cs}}\right) - \nonumber \\
&-& \ln\Bigg[\cosh\Bigg(\frac{x}{w_\mathrm{cs}}\Bigg)\Bigg] \mathrm{sech}^2\Bigg(\frac{x}{w_\mathrm{cs}}\Bigg)\Bigg\}\exp\Bigg(-\frac{2y}{\lambda}\Bigg), \nonumber \\
\end{eqnarray}
where
\begin{equation}\label{eq16}
p_h(y) = p_0 \exp\left[-\int\limits_{y_0}^{y} \frac{1}{\Lambda(\tilde{y})}\mathrm{d}\tilde{y}\right],
\end{equation}
and
\begin{equation}\label{eq17}
\varrho_h(y) = \frac{p(y)}{g_{\sun}\Lambda(y)}.
\end{equation}
Here
\begin{equation}\label{eq18}
\Lambda(y) = \frac{k_\mathrm{B}T(y)}{\overline{m}g_{\sun}}
\end{equation}
is the pressure scale-height which in the case of isothermal atmosphere represents the vertical distance over which the gas pressure decreases by a factor of $e\approx2.7$, $k_\mathrm{B} = 1.38 \times 10^{-23}~\mathrm{J\cdot K^{-1}}$ is the Boltzmann constant and $\overline{m} = 0.6\, m_\mathrm{p}$ is the mean particle mass ($m_\mathrm{p} = 1.672 \times 10^{-27}~\mathrm{kg}$ is the proton mass). $p_0$ in Eq. (\ref{eq16}) denotes the gas pressure at the reference level $y_0$. In our calculations we set $y_0 = 10~\mathrm{Mm}$.

For the solar atmosphere, the temperature profile $T(y)$, was derived by~\citet{1981ApJS...45..635V, 2008ApJS..175..229A}. At the top of the photosphere, which corresponds to the height of $y = 0.5~\mathrm{Mm}$, the temperature is $T(y) = 5700~\mathrm{K}$. At higher altitudes, the temperature falls down to its minimal value $T(y) = 4350~\mathrm{K}$ at $y \approx 0.95~\mathrm{Mm}$. Higher up the temperature rises slowly to the height of about $y = 2.7~\mathrm{Mm}$,  where the transition region (TR) is located. Here the temperature increases abruptly to the value $T(y) = 1.5~\mathrm{MK}$, at the altitude $y = 10~\mathrm{Mm}$, which is typical for the solar corona.
\subsection{Perturbations}
At the beginning of the numerical simulation ($t = 0~\mathrm{s}$), the equilibrium is perturbed by a Gaussian compression pulse in the $x$-component of velocity and has the following form \citep[e.g.][]{2004MNRAS.349..705N,2005SSRv..121..115N}:

\begin{equation}\label{eq19}
v_x = -A_0 \frac{x}{\lambda_y} \exp{\left[-\left(\frac{x}{\lambda_x}\right)^2\right]}  \exp{\left[-\left(\frac{y-L_\mathrm{P}}{\lambda_y}\right)^2\right]},
\end{equation}
where $A_0\mathbf{=1 \mathrm{km}\cdot{\mathrm{s}^{-1}}}$ is the initial amplitude of the pulse, and $\lambda_x = \lambda_y = 0.15~\mathrm{Mm}$ are the widths of the velocity pulse in the longitudinal and transverse directions, respectively. For such an initial velocity pulse, the ratio between the maximum of kinetic energy density and the background (thermal) energy density is $\approx 4.167 \times 10^{-5}$. The perturbation point $(x=0,y=L_\mathrm{P})$ is located on the axis of the current-sheet, at a distance of $L=25~\mathrm{Mm}$ from the bottom boundary of the simulation region. This pulse is used in order to start the primary tearing instability in the current-sheet. Later on, because of the resistivity used in our model, a number of plasmoids as well as the magnetic arcade in the bottom part of the simulation region is formed self-consistently.

\section{Numerical solutions and results}
We solve the 2-D time-dependent, ideal MHD equations (1)-(4) numerically, making use of the FLASH code \citep{2009JCoPh.228..952L,2013JCoPh.243..269L}. It is a well tested, fully modular, parallel, multi-physics, open science, simulation code that implements second- and third-order unsplit Godunov solvers with various slope limiters and Riemann solvers as well as adaptive mesh refinement \citep[AMR, e.g.][]{2002cfd..book.....C}. The Godunov solver combines the corner transport upwind method for multi-dimensional integration and the constrained transport algorithm for preserving the divergence free constraint on the magnetic field \citep{2009JCoPh.228..952L}. We use the minmod slope limiter and the Riemann solver \citep[e.g.][]{2006IJNMF..52..433T}. The main advantage of using the AMR technique is to refine a numerical grid at steep spatial profiles while keeping the grid coarse at the places where fine spatial resolution is not essential. In our case, the AMR strategy is based on controlling the numerical errors near the gradient of mass density, leading to a reduction of the numerical diffusion within the entire simulation region.

\begin{figure*}[h!]
\centering
\includegraphics[scale=0.45]{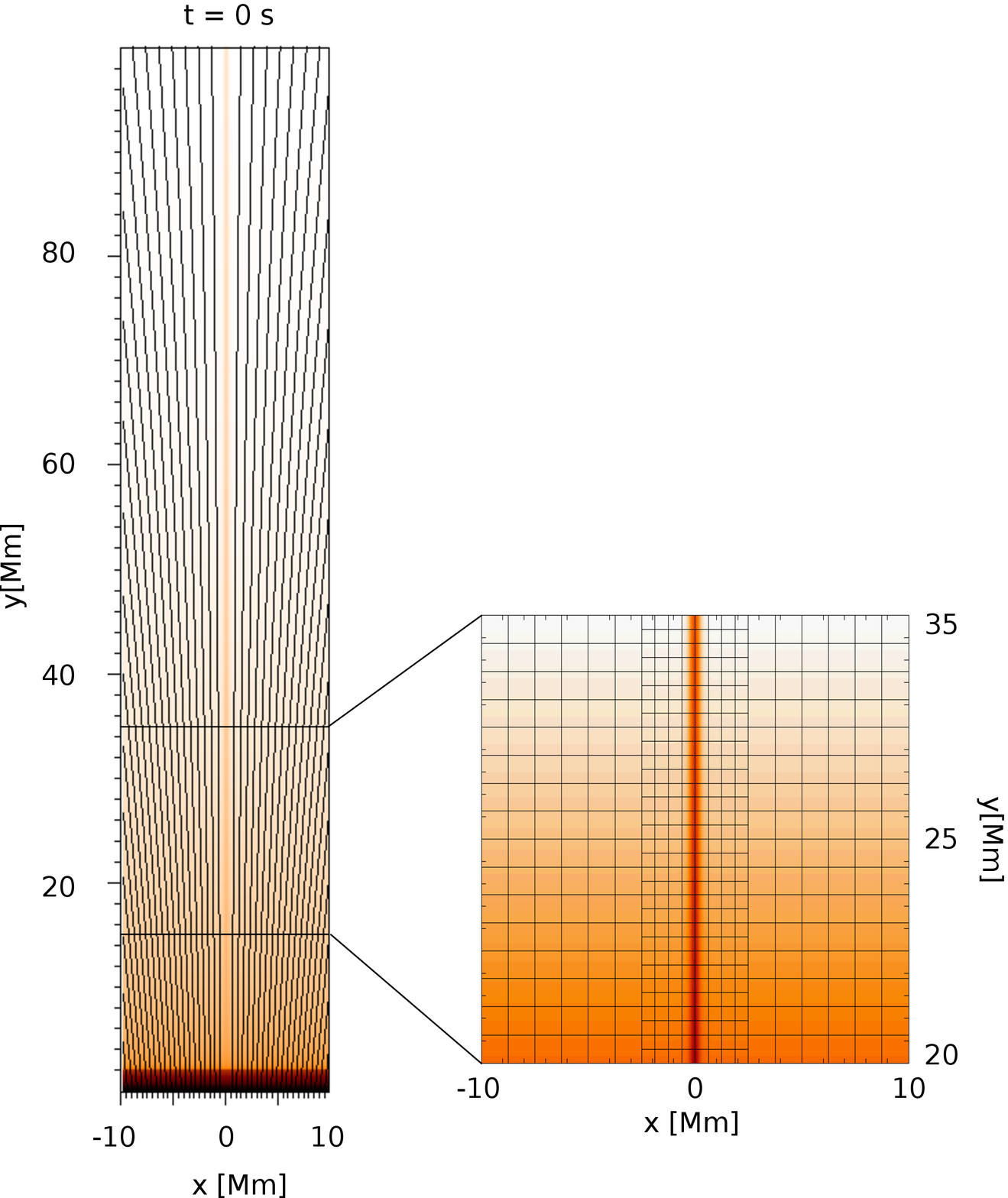}
\caption{Left: Initial mass density distribution over-plotted by the magnetic field structure as black lines. Right: The detail of the simulation region in the vicinity of perturbation point $(25~\mathrm{Mm})$, illustrating the computational grid, with adaptive mesh refinement (AMR).} \label{Fig1}
\end{figure*}

For our numerical simulations, we use a 2-D Eulerian box of height $H = 100~\mathrm{Mm}$ and width $W = 20~\mathrm{Mm}$. The spatial resolution of the numerical grid is determined with the AMR method and we use a similar setup as in the case of the vertical current-sheet presented in \citet{2015A&A...581A.131J}. We use an AMR grid with the minimum (maximum) level of the refinement blocks set to $3 (6)$ to have $203~328$ numerical cells (Fig. \ref{Fig1}).

Prior to performing the numerical simulations, we verified that the system remains in numerical equilibrium for the adopted grid resolution, by running a test simulation without any velocity pulse.

Very shortly after the initial pulse, the magnetic field reconnects and plasmoids (2-D O-type magnetic field structures) are generated. The plasmoids move up and down in the current sheet, exciting oscillations in the plasmoids themselves and the magnetic arcade. These processes are described in the following subsections. For better visibility of oscillations and other processes we also enclosed the movie from our numerical simulations.

\subsection{Plasmoid merging and oscillations}
\begin{figure*}[h!]
\centering
\includegraphics[scale=0.45]{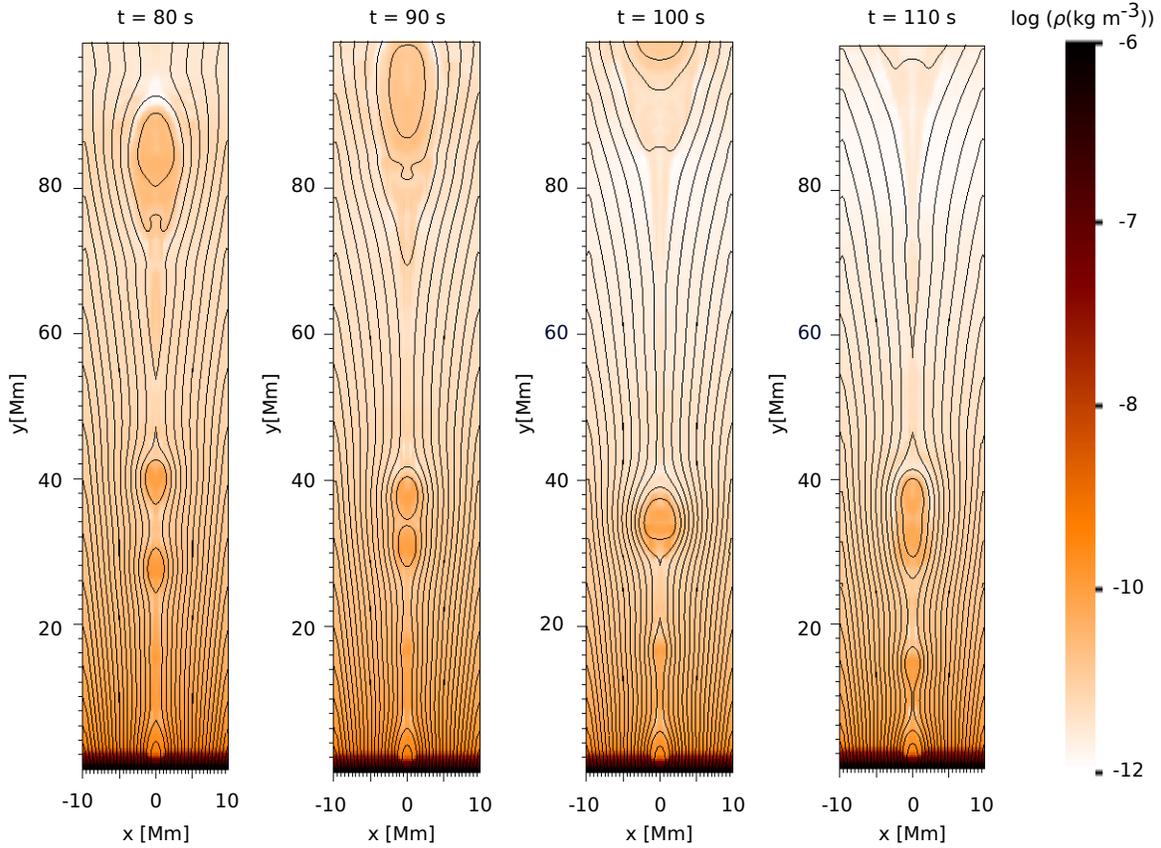}
\caption{The mass density and corresponding magnetic field lines, illustrating the merging of two plasmoids followed by an oscillation of the resulting plasmoid (see the processes in the $y = 20 - 50~\mathrm{Mm}$ interval). The panels from left to right show times $t=80, 90, 100$ and $110~\mathrm{s}$, respectively.} \label{Fig2}
\end{figure*}

In Fig. \ref{Fig2}, we show the mass density and magnetic field lines for times $80,90,100$ and $110~\mathrm{s}$, respectively. After the beginning of the magnetic reconnection, a number of plasmoids are formed in the vertical current sheet due to the tearing mode instability. During the simulation, some plasmoids merge into bigger ones and move upwards through the corona and downwards to the solar photosphere, because of the gravitational or buoyancy forces. At time $t=80~\mathrm{s}$ and in the $y = 20-50~\mathrm{Mm}$ interval (Fig. \ref{Fig2}, first panel), there are two plasmoids which move towards each other. Around time $t=90~\mathrm{s}$ (Fig. \ref{Fig2}, second panel), they start to merge into one larger plasmoid which then oscillates. These oscillations are clearly visible at times $100$ and $110~\mathrm{s}$. At that time, the oscillating plasmoid inverts its sense of propagation direction and starts to move up, see the enclosed movie. At the time around $160~\mathrm{s}$, the oscillations are almost completely damped and the plasmoid slowly increases its size (in height $70-80~\mathrm{Mm}$) and at time around $200~\mathrm{s}$ leaves freely the simulation region, as can also be seen in the movie.

\begin{figure*}[h!]
 \hspace{-0.5cm}
 \begin{center}
  \includegraphics[scale = 0.3]{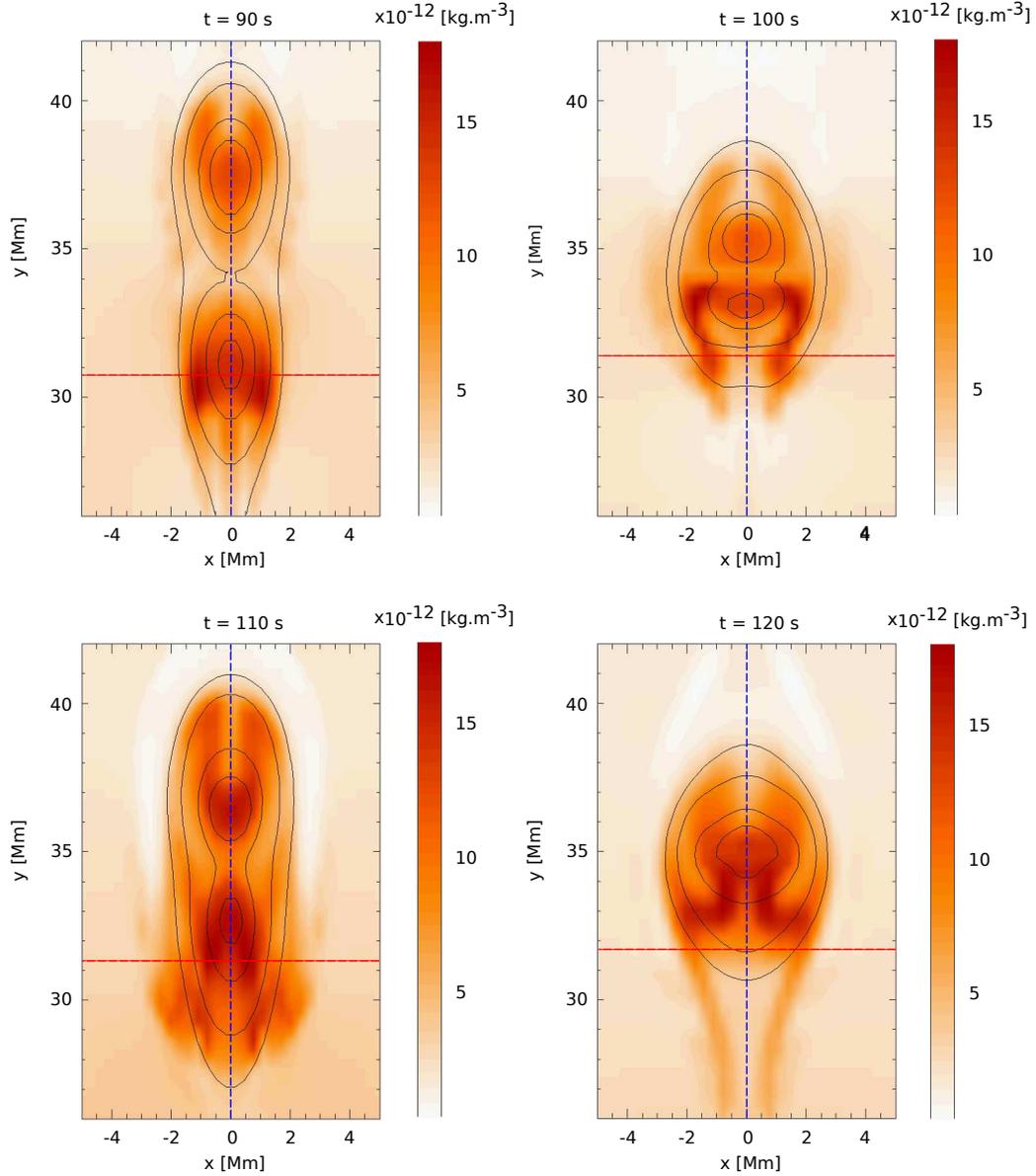}
 \caption{A detailed view of the mass density and magnetic field lines of the coalescing plasmoids at different times ($t=90, 100, 110, 120~\mathrm{s}$), showing their self-oscillations and complex density structure.}
 \label{Fig3}
 \end{center}
\end{figure*}

Fig. \ref{Fig3} provides a zoom-in for a detailed look at the behaviour of the oscillating plasmoid and its inner structure for times $90,100,110$ and $120~\mathrm{s}$. The blue and red dashed lines correspond to a 1D density slice for both, a vertical direction and a horizontal direction, respectively. The positions of slices are chosen on the axis of symmetry of simulation region $(x=0.0~\mathrm{Mm})$, panels marked as (a), and at the place, where we found the maximum of mass density on this axis, panels marked as (b) -- ($\approx 32-33~\mathrm{Mm}$), respectively -- see Fig. \ref{Fig4}. Two peaks in graphs (a) for times $80-100~\mathrm{s}$ represent two plasmoids which move towards each other and merge into one plasmoid which is finally represented by an increase of mass density in one peak visible at time $110~\mathrm{s}$. In all displayed panels (b) we can recognize very nicely the complex internal plasmoid structure shortly before and after the merging process.

Figure \ref{Fig5} shows the interaction of the two plasmoids followed by oscillations of the merged larger plasmoid. Essentially, the figure shows the temporal map of the vector potential at heights between $18-50~\mathrm{Mm}$. To show clearly the plasmoids and their merging in this color map we use only the selected values of the vector potential corresponding to their magnetic field lines. Here, we can see very clearly the plasmoid motion into higher altitudes of the solar atmosphere at times after $125~\mathrm{s}$. The plasmoid $(\mathrm{P}_1)$ and plasmoid $(\mathrm{P}_2)$ are determined by two magnetic field lines (black solid lines) on the current sheet axis determined by the vector potentials $A = -2.45$ and $A = -4.9$. P indicates the resulting plasmoid which starts to oscillate with a $\tau_\mathrm{plas} \approx 25~\mathrm{s}$ period, after the merger from the two plasmoids $(\mathrm{P}_1)$ and $(\mathrm{P}_2)$. This period is very close to that one calculated from the formula $\tau_\mathrm{plas} = L / c_\mathrm{A} = 27.9~\mathrm{s}$, where $L \approx 6~\mathrm{Mm}$ is the size of plasmoid and $c_\mathrm{A} \approx 0.215~\mathrm{Mm}\cdot\mathrm{s^{-1}}$ is the Alfv\'{e}n speed in the lateral parts of the plasmoid \citep{1987ApJ...321.1031T}.

To show clearly the oscillatory process in the newly formed bigger plasmoid we
present Fig. \ref{Fig6}. Here we can see the temporal evolution of the maximum
of the mass density on the axis of current sheet between the heights
$25-42~\mathrm{Mm}$, where a clear oscillatory behaviour of this plasmoid is
observed. We can see here a wavy signal with an initial rarefaction around the
time $\approx 90~\mathrm{s}$ followed by a compression at time $\approx
95~\mathrm{s}$. This initial phase continues with the next two oscillations
which are very quickly attenuated, so that at time $\approx 160~\mathrm{s}$
they are damped entirely. Firstly, we checked if this strong damping of
oscillations is caused by the magnetic diffusivity. But, the diffusion time
calculated as $\tau_\mathrm{damp} \approx \mathcal{L}^2/\eta = 2.5 \times
10^9~\mathrm{s}$, see \citep{1967imhd.book.....R}, where $\mathcal{L}\approx
5~\mathrm{Mm}$ is the plasmoid width and the assumed magnetic diffusivity
$\eta\approx100~\mathrm{m^2 \cdot s^{-1}}$, is very long. Secondly, we
performed a calculation with twice higher spatial resolution and find that the
damping time essentially was not changed. Furthermore, we made a special MHD
test with a simple oscillating slab and we varied the resistivity in the
interval from zero up to the resistivity, which is 100 times greater than the
collisional resistivity (the resistivity used in this paper). In this test, in
the mentioned interval of the resistivity, no change of the damping was found.
Therefore, we made a detailed analysis and we found that the plasmoid is not an
isolated oscillating system, it is strongly coupled with plasma flows in its
vicinity, and the damping (decrease of the oscillation amplitude) is caused by
these mass plasma flows, see the enclosed movie, showing the plasma velocity
field.

\subsection{Arcade oscillations}
Fig. \ref{Fig7} shows the mass density and magnetic field lines, for the times during the collision of the plasmoid with the magnetic arcade. At time $t=120~\mathrm{s}$ and for $y$ below $20~\mathrm{Mm}$ (Fig. \ref{Fig7}, first panel), the plasmoid moves downwards to the magnetic arcade because of the gravitational force and tension of the surrounding magnetic field lines. Shortly after, at time $t=130~\mathrm{s}$ (Fig. \ref{Fig7}, second panel), the plasmoid and arcade merge into one bigger arcade which starts to oscillate, as can be seen at times $t=140$ and $150~\mathrm{s}$ (Fig. \ref{Fig7}, third and last panel), see also the enclosed movie. From our numerical simulations we found that the arcade oscillations are also very quickly damped. The reason for this damping is exactly the same as in the case of plasmoids oscillations. The arcade is not an isolated system, it is strongly coupled with processes in the current sheet. The downward oriented plasma flow, carrying plasmoids from reconnection towards the arcade, not only triggers the arcade oscillation, but in the following times the mass plasma flow damps this arcade oscillation, see the enclosed movie.

\begin{figure*}[h!]
\begin{center}
\footnotesize{$t=90~\mathrm{s}$}
\end{center}
\begin{center}
\vspace*{-0.3cm}
\hspace*{-0.5cm}
\mbox{
\includegraphics[scale = 0.27]{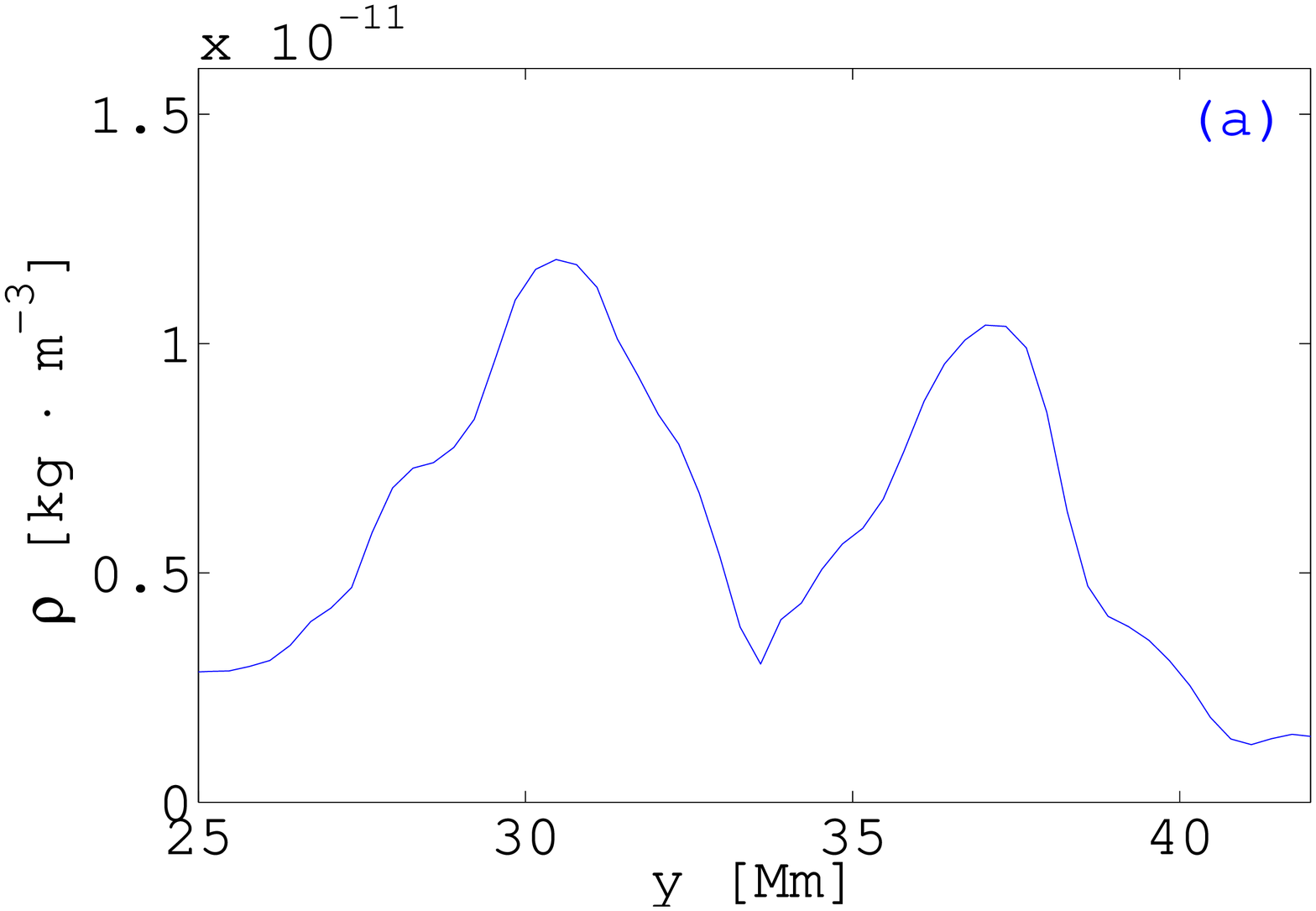}
\includegraphics[scale = 0.27]{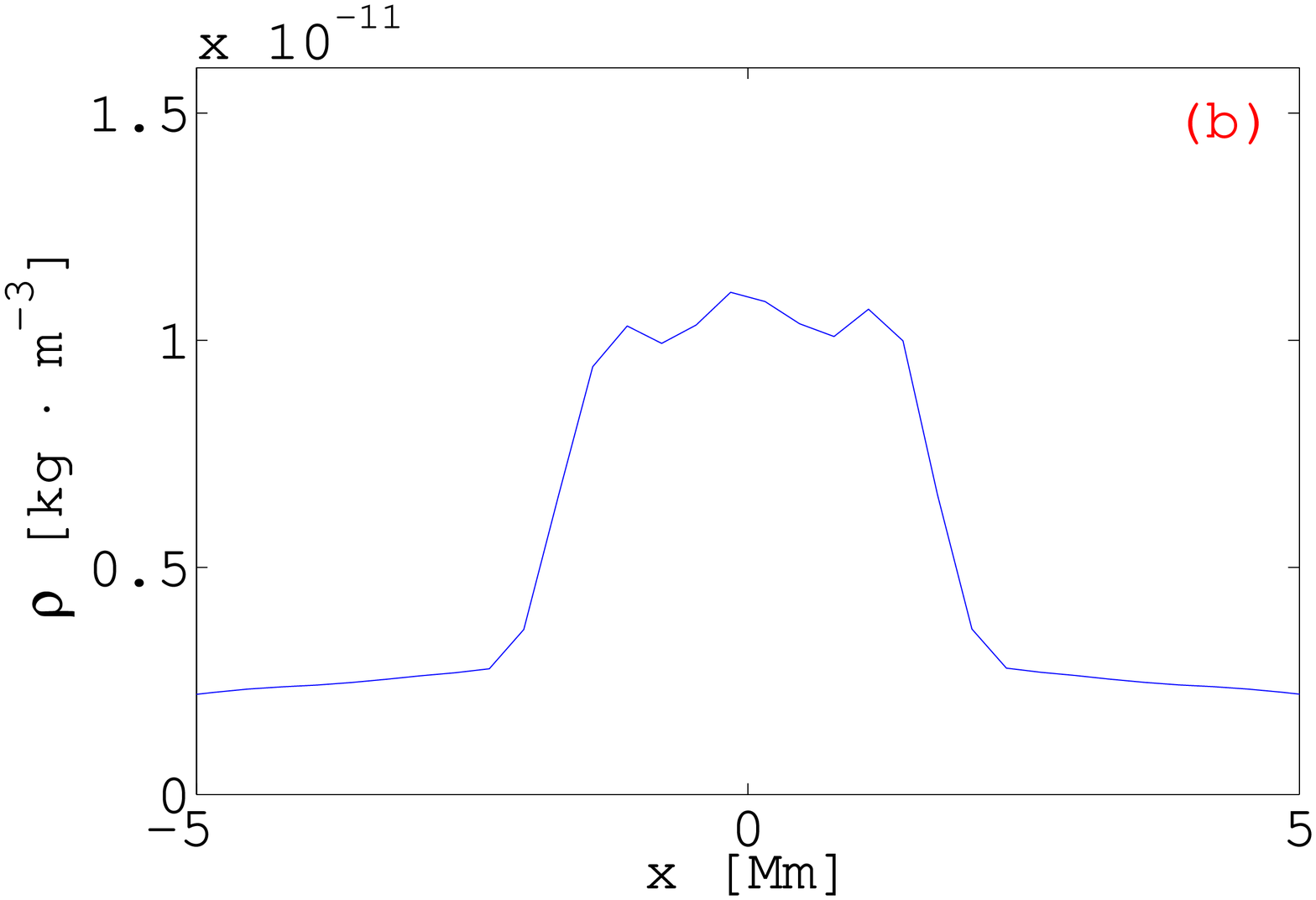}
}
\vspace*{-0.5cm}
\end{center}
\begin{center}
\footnotesize{$t=100~\mathrm{s}$}
\end{center}
\begin{center}
\vspace*{-0.3cm}
\hspace*{-0.5cm}
\mbox{
\includegraphics[scale = 0.27]{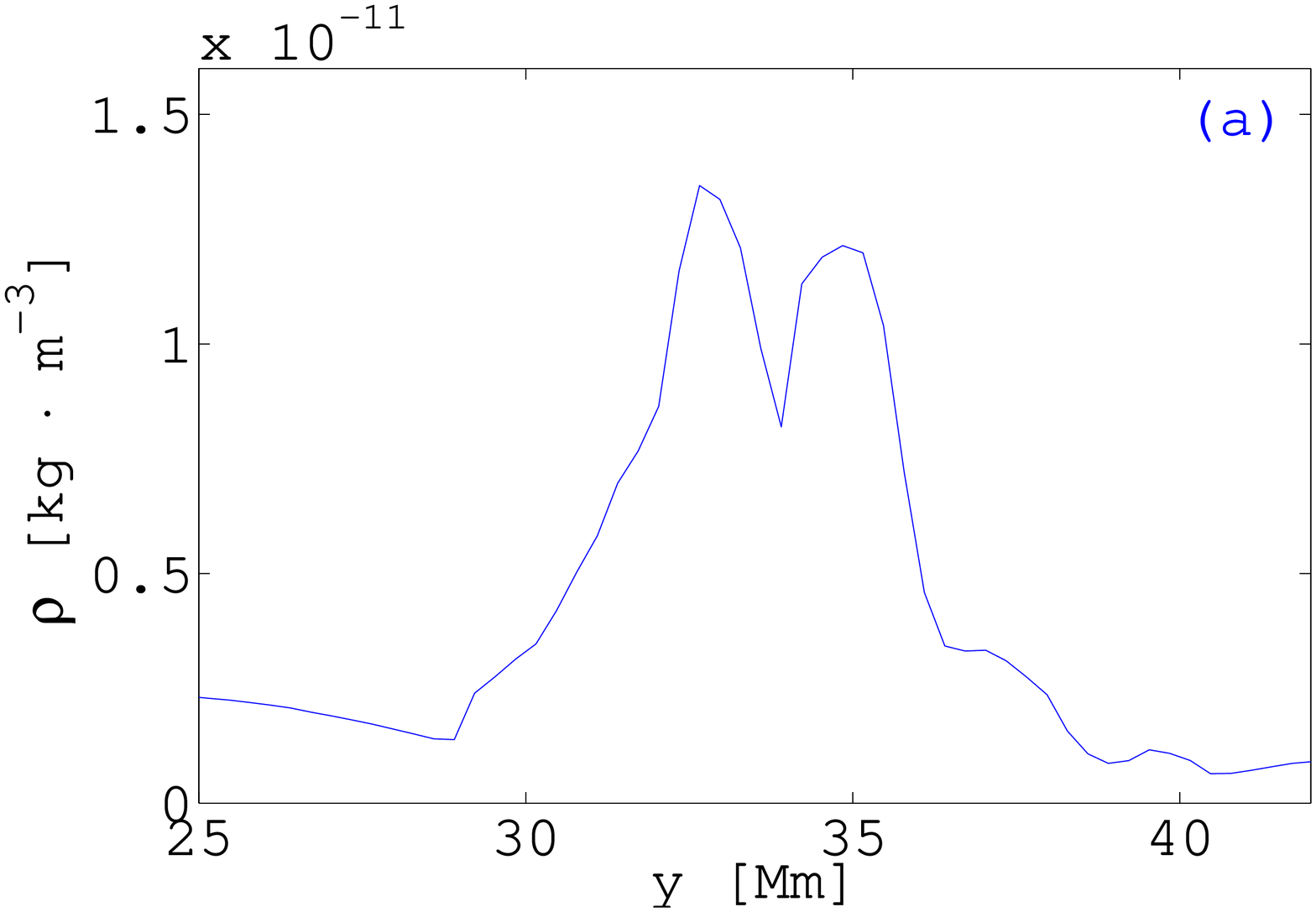}
\includegraphics[scale = 0.27]{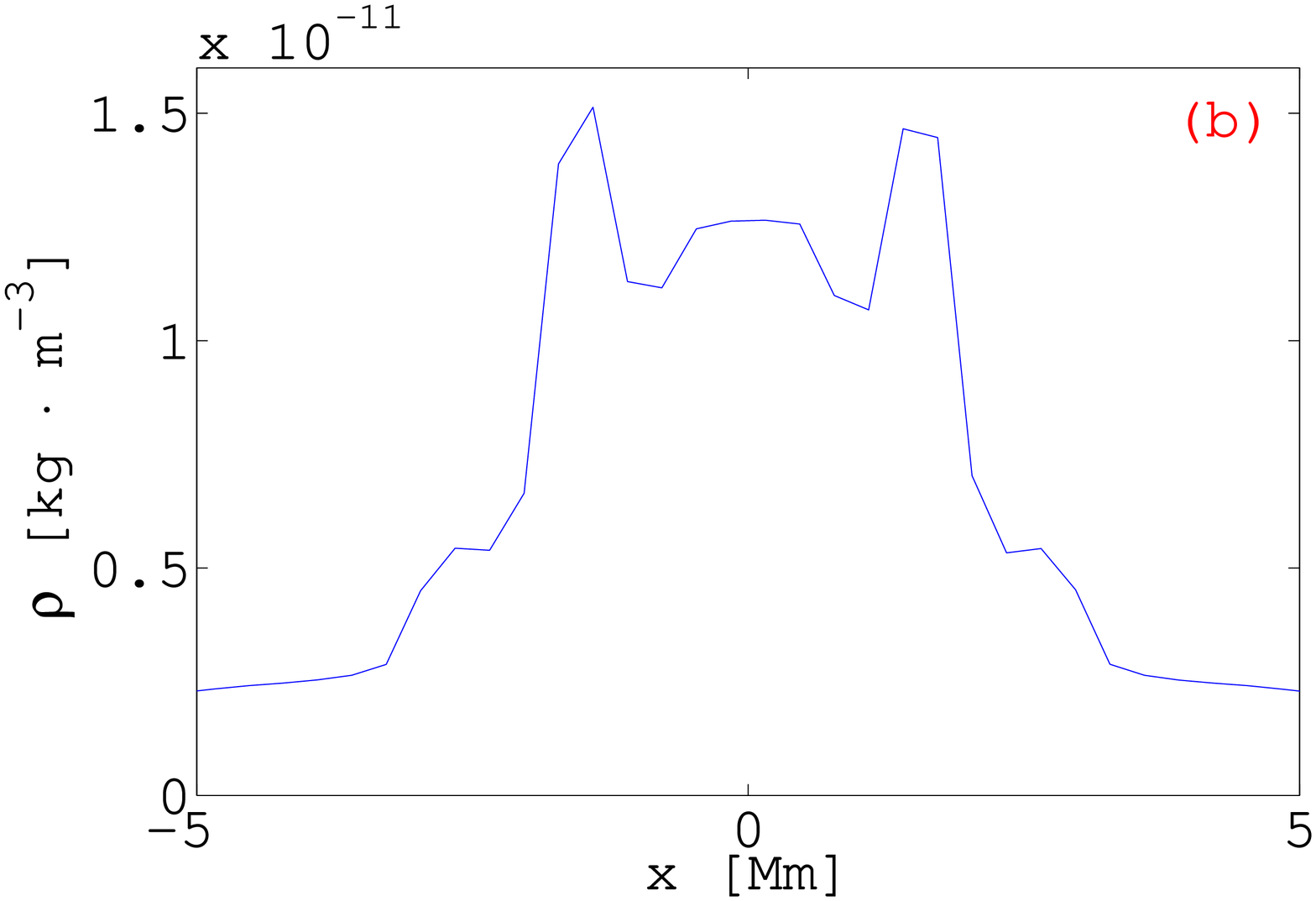}
}
\vspace*{-0.5cm}
\end{center}
\begin{center}
\footnotesize{$t=110~\mathrm{s}$}
\end{center}
\begin{center}
\vspace*{-0.3cm}
\hspace*{-0.5cm}
\mbox{
\includegraphics[scale = 0.27]{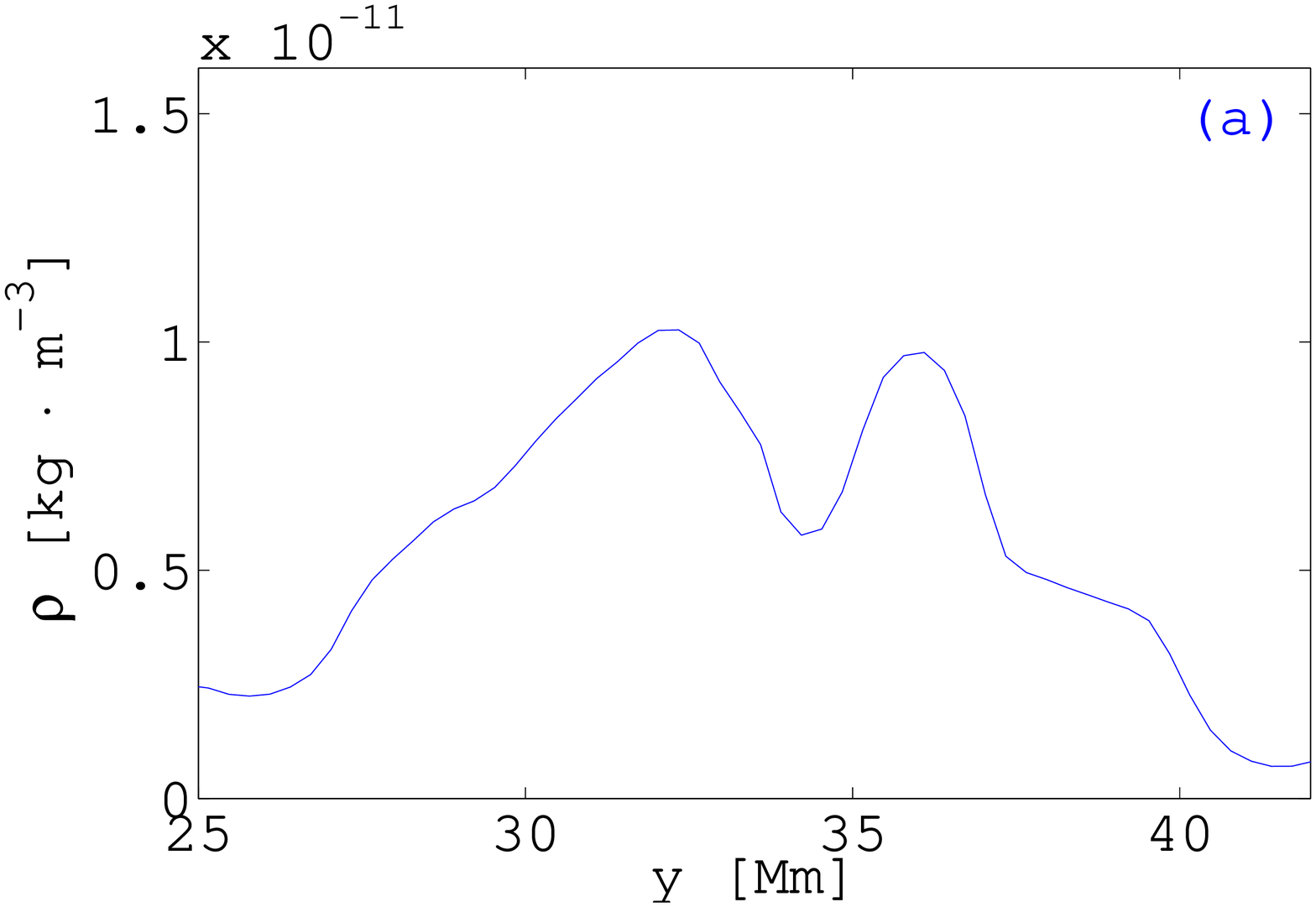}
\includegraphics[scale = 0.27]{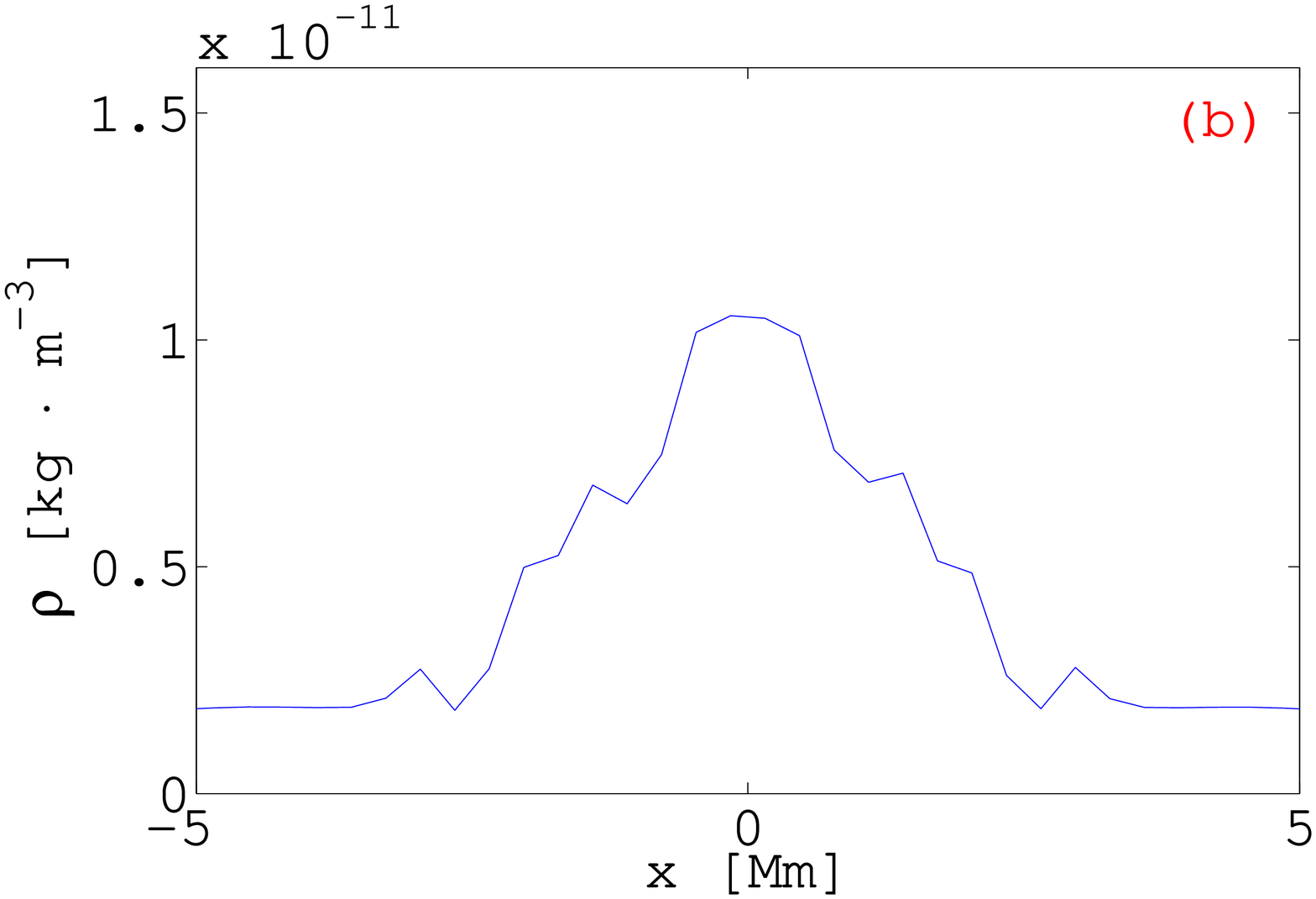}
}
\vspace*{-0.5cm}
\end{center}
\begin{center}
\footnotesize{$t=120~\mathrm{s}$}
\end{center}
\begin{center}
\vspace*{-0.3cm}
\hspace*{-0.5cm}
\mbox{
\includegraphics[scale = 0.27]{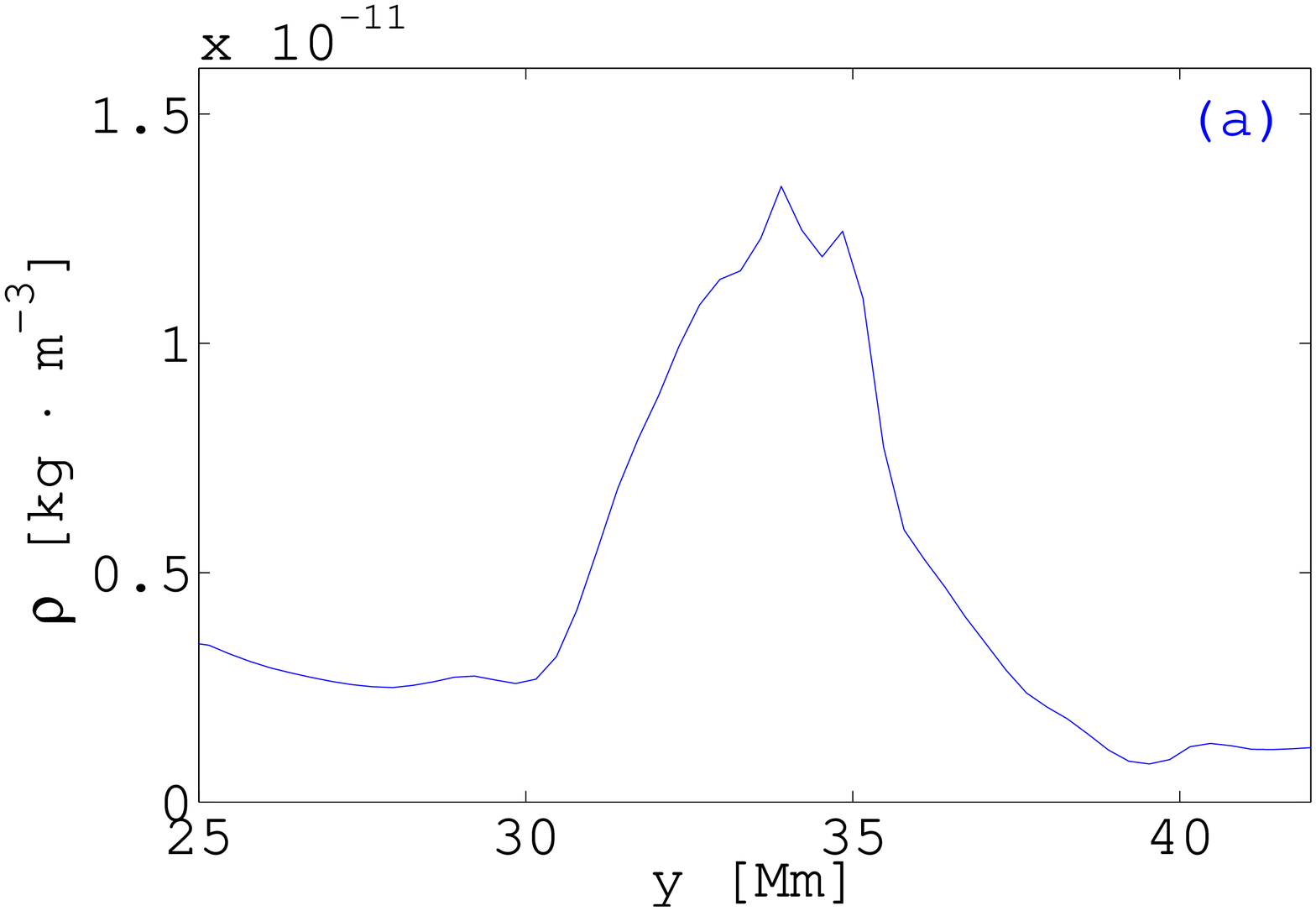}
\includegraphics[scale = 0.27]{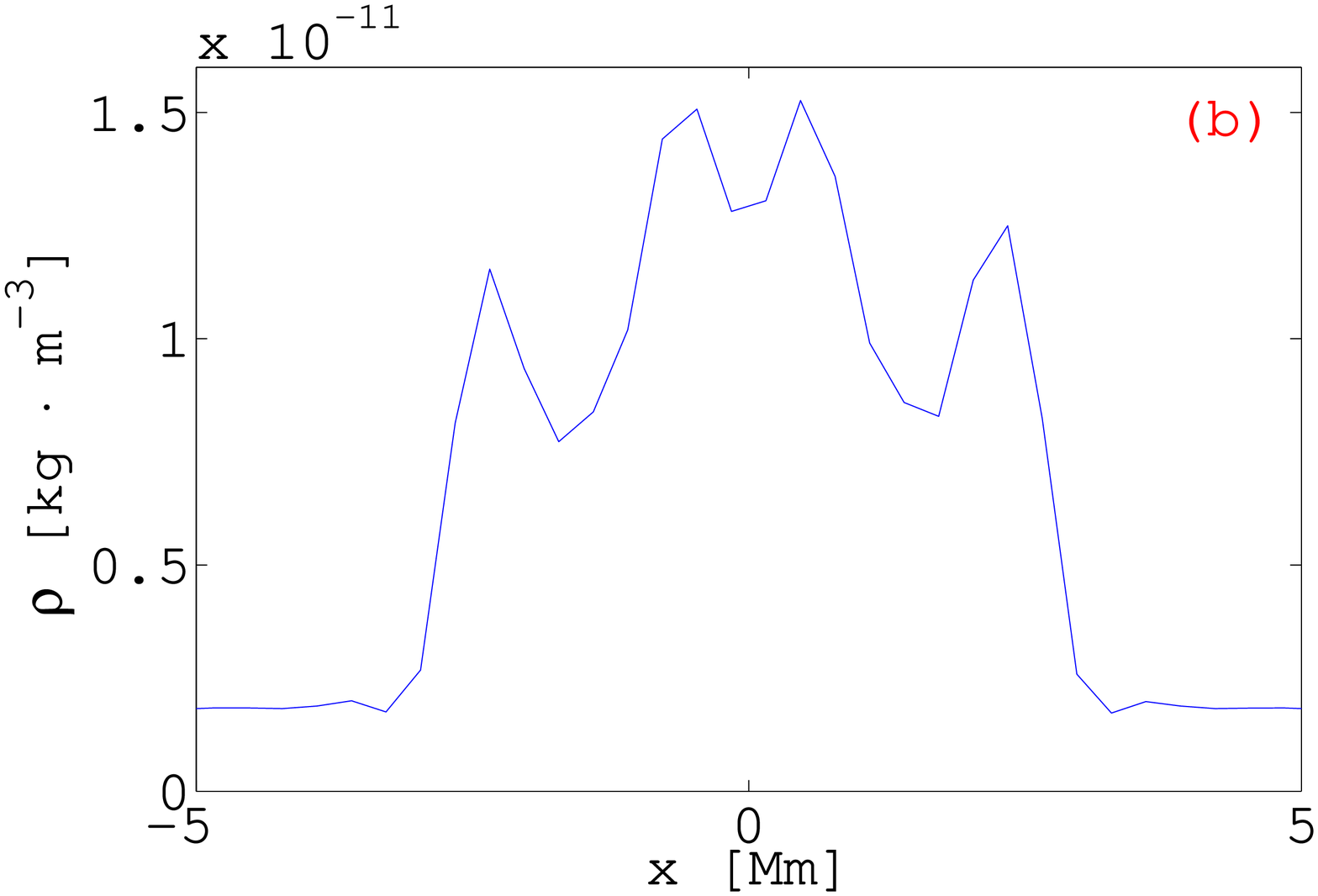}
}
\vspace*{-0.5cm}
\end{center}
\caption{Density profiles along the vertical blue (a) and horizontal red (b) lines in Fig.~\ref{Fig3}.} \label{Fig4}
\end{figure*}

\begin{figure}[h!]
 \hspace{-0.5cm}
\includegraphics[scale = 0.23]{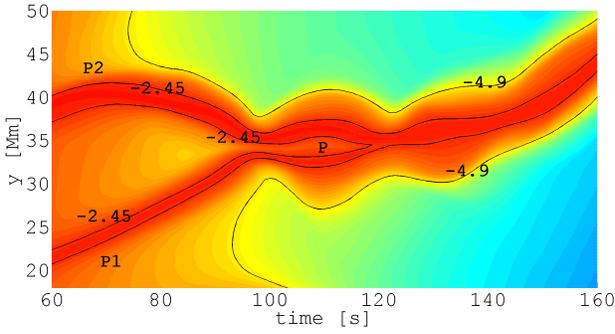}
\caption{The temporal map of the vector potential at heights $18-50~\mathrm{Mm}$ above the solar surface. The solid black lines represent the positions of the magnetic field lines with the vector-potential $A=-2.45$ and $A=-4.9$ along the axis of the current sheet during the merging of the two plasmoids (P1 and P2) into one plasmoid (P).}
\label{Fig5}
\end{figure}

\begin{figure}[h!]
 \hspace{-0.5cm}
 \begin{center}
 \includegraphics[scale = 0.24]{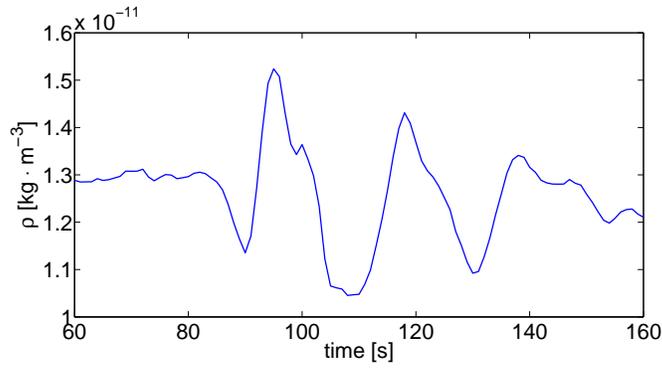}
 \caption{The evolution of the maximum mass density in the merging plasmoids, showing periodic behaviour with a period $\approx 25~\mathrm{s}$.}
 \label{Fig6}
 \end{center}
\end{figure}

\begin{figure*}[h!]
\centering
\includegraphics[scale=0.45]{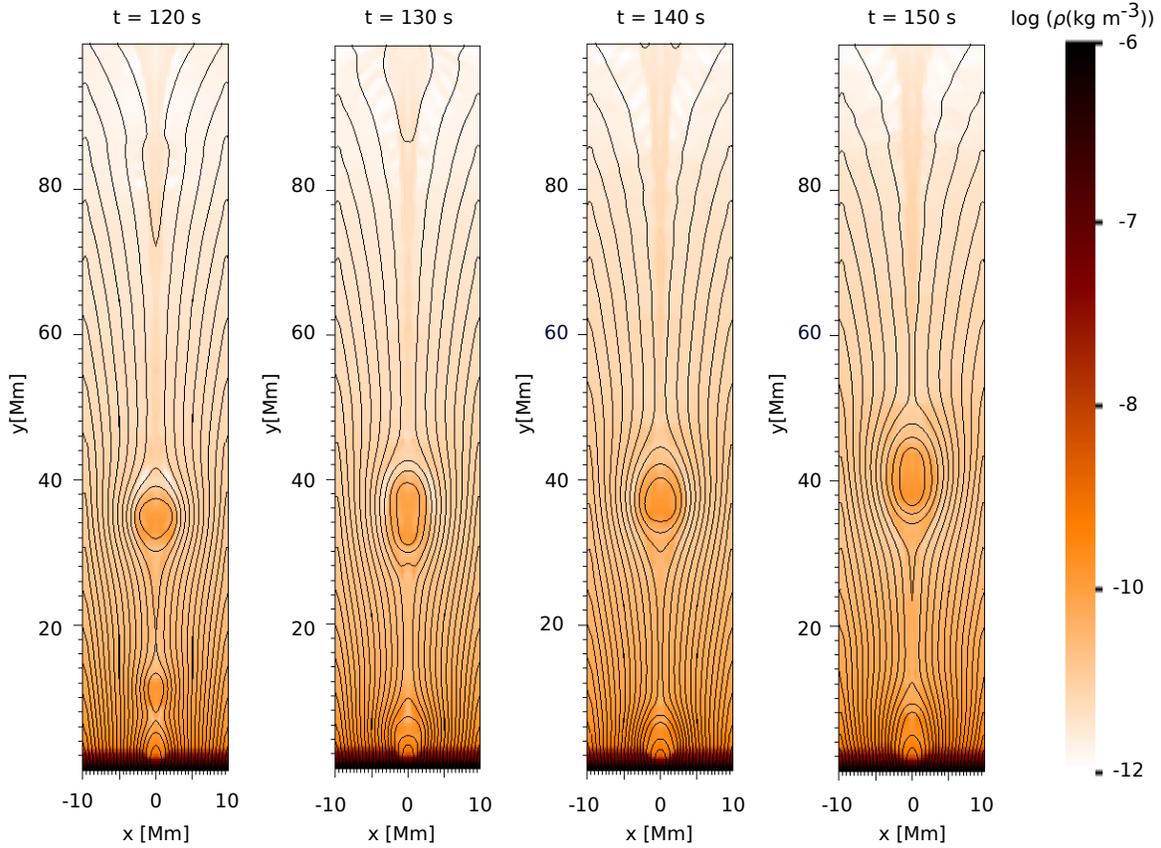}
\caption{The mass density and corresponding magnetic field lines, illustrating the collision of the plasmoid with the magnetic arcade, showing its quickly damped oscillation (see the processes below $y = 20~\mathrm{Mm}$). The panels from left to right show times $t=120, 130, 140$ and $150~\mathrm{s}$, respectively.}
\label{Fig7}
\end{figure*}

\begin{figure*}[h!]
 \hspace{-0.5cm}
 \begin{center}
  \includegraphics[scale = 0.3]{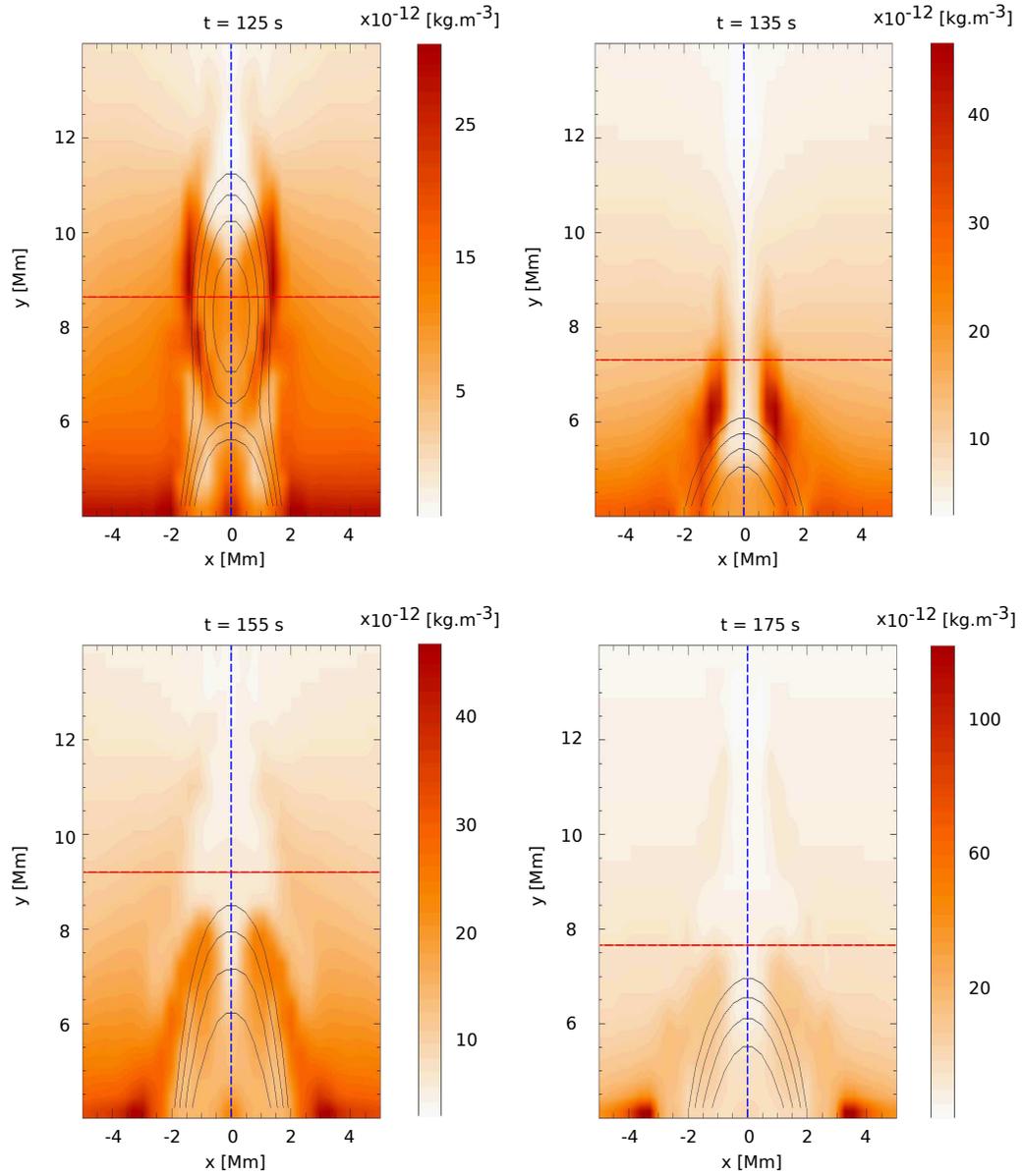}
 \caption{A detailed view of densities and magnetic field lines during the merging of the plasmoid with the arcade at times $t=125, 135, 155, 175~\mathrm{s}$.}
 \label{Fig8}
 \end{center}
\end{figure*}

\begin{figure*}[h!]
\begin{center}
\footnotesize{$t=125~\mathrm{s}$}
\end{center}
\begin{center}
\vspace*{-0.3cm}
\hspace*{-0.5cm}
\mbox{
\includegraphics[scale = 0.27]{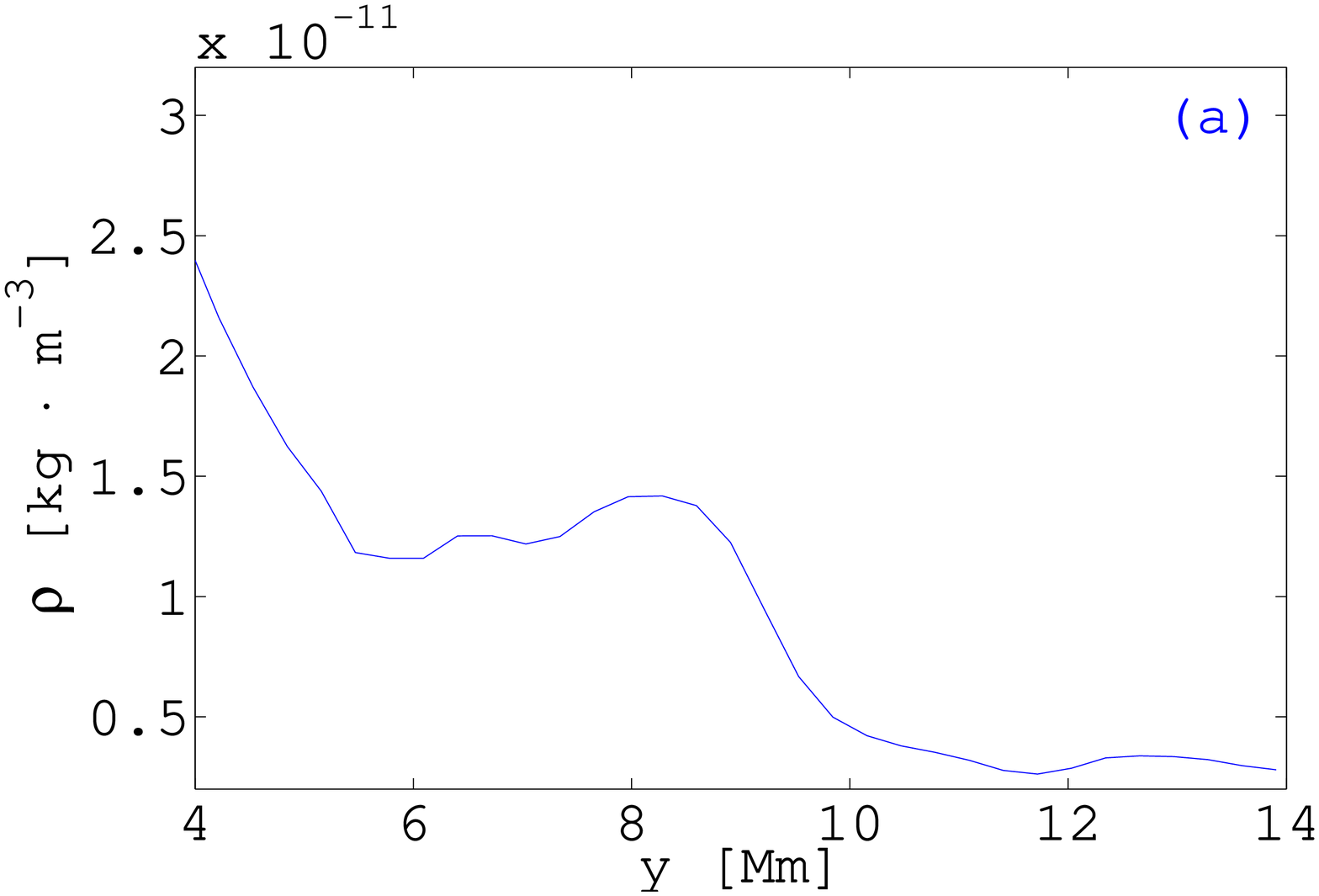}
\includegraphics[scale = 0.27]{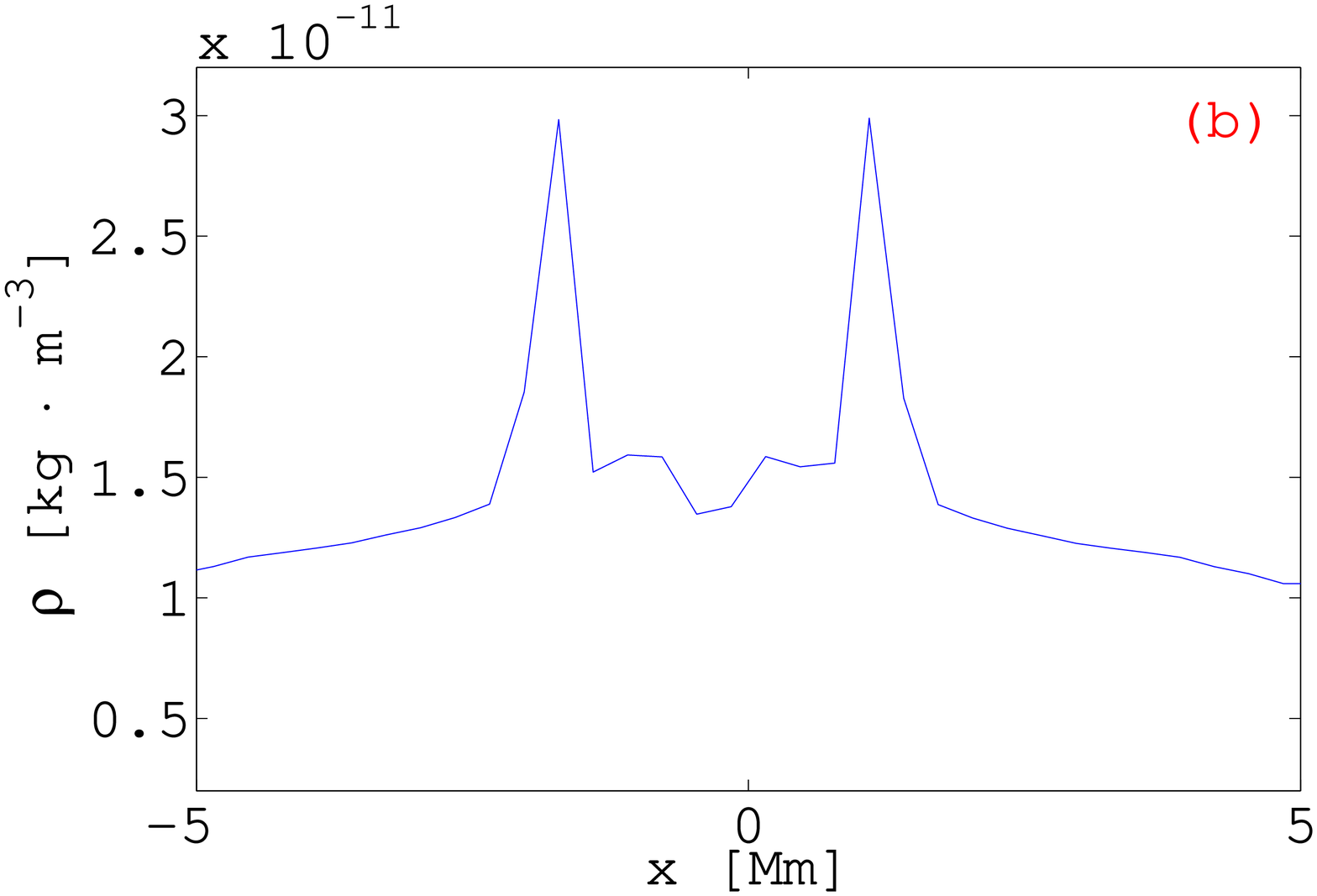}
}
\vspace*{-0.5cm}
\end{center}
\begin{center}
\footnotesize{$t=135~\mathrm{s}$}
\end{center}
\begin{center}
\vspace*{-0.3cm}
\hspace*{-0.5cm}
\mbox{
\includegraphics[scale = 0.27]{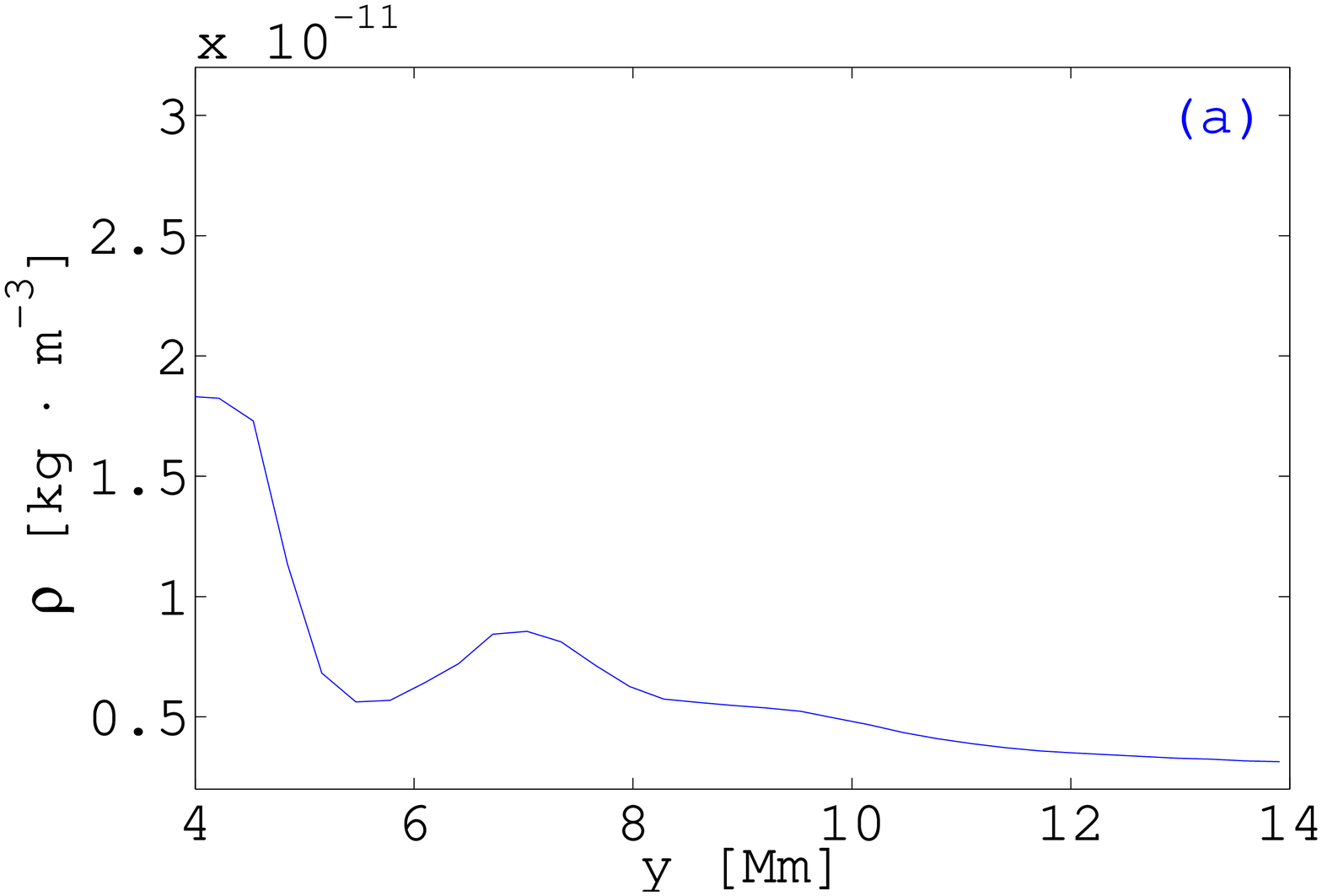}
\includegraphics[scale = 0.27]{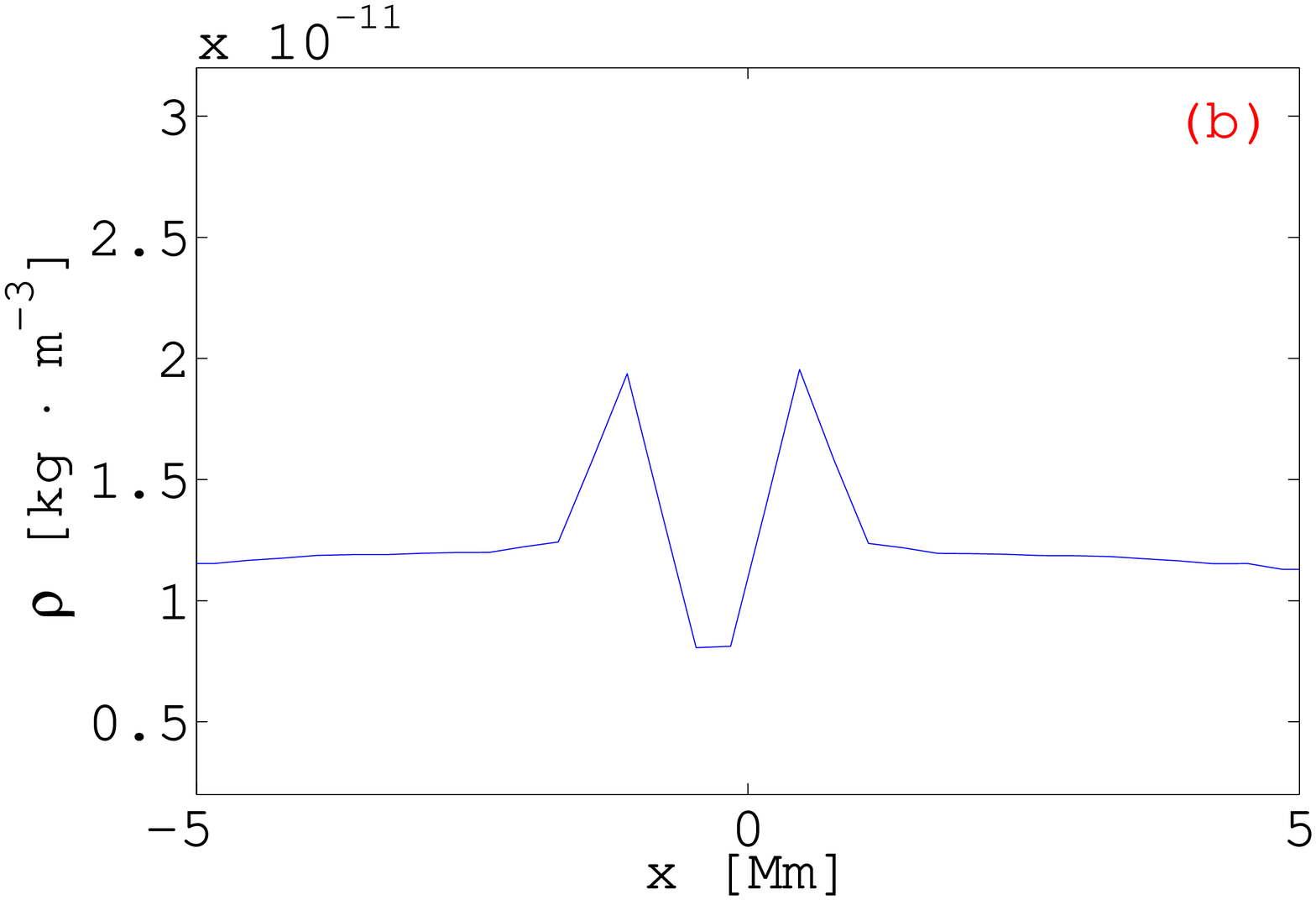}
}
\vspace*{-0.5cm}
\end{center}
\begin{center}
\footnotesize{$t=155~\mathrm{s}$}
\end{center}
\begin{center}
\vspace*{-0.3cm}
\hspace*{-0.5cm}
\mbox{
\includegraphics[scale = 0.27]{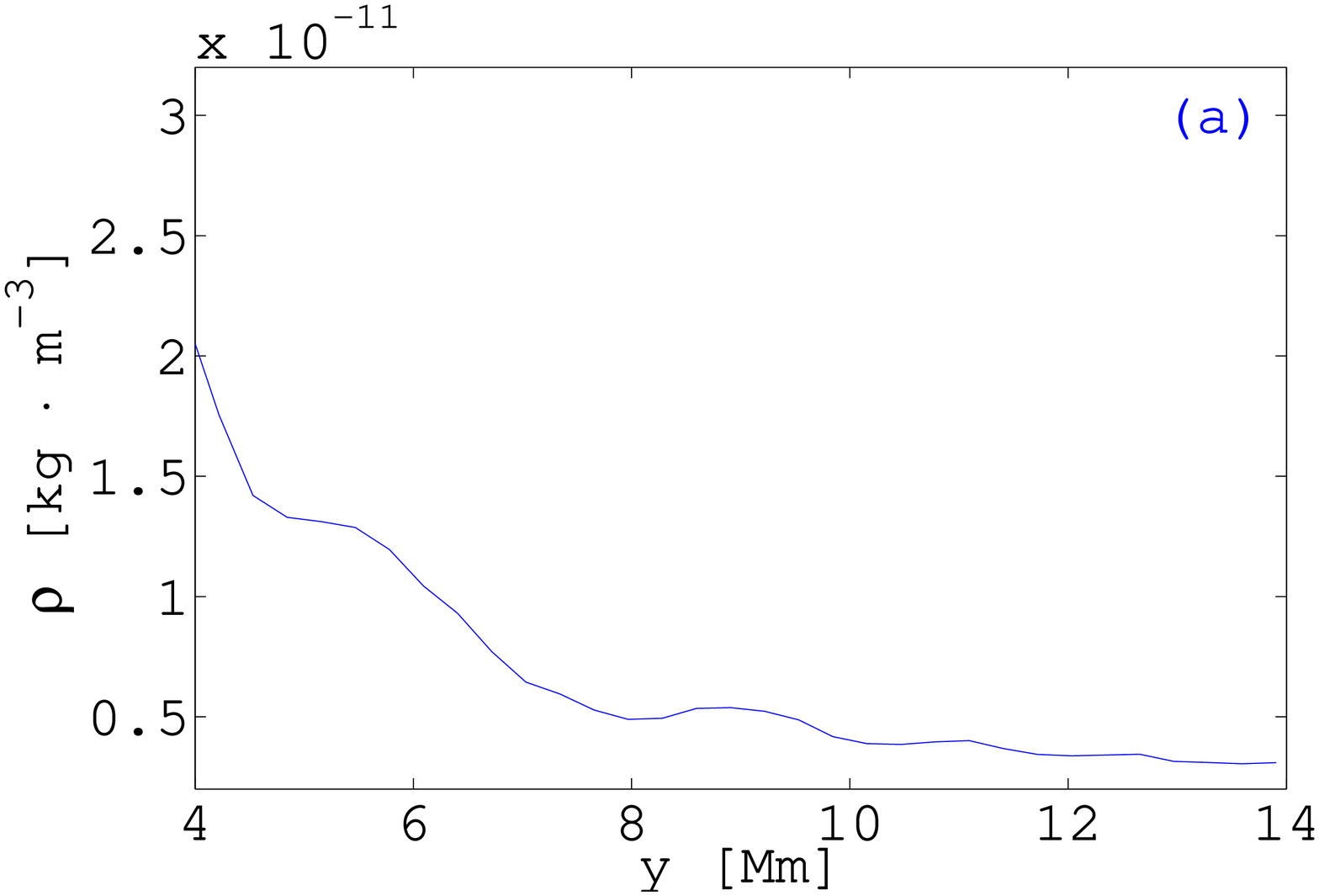}
\includegraphics[scale = 0.27]{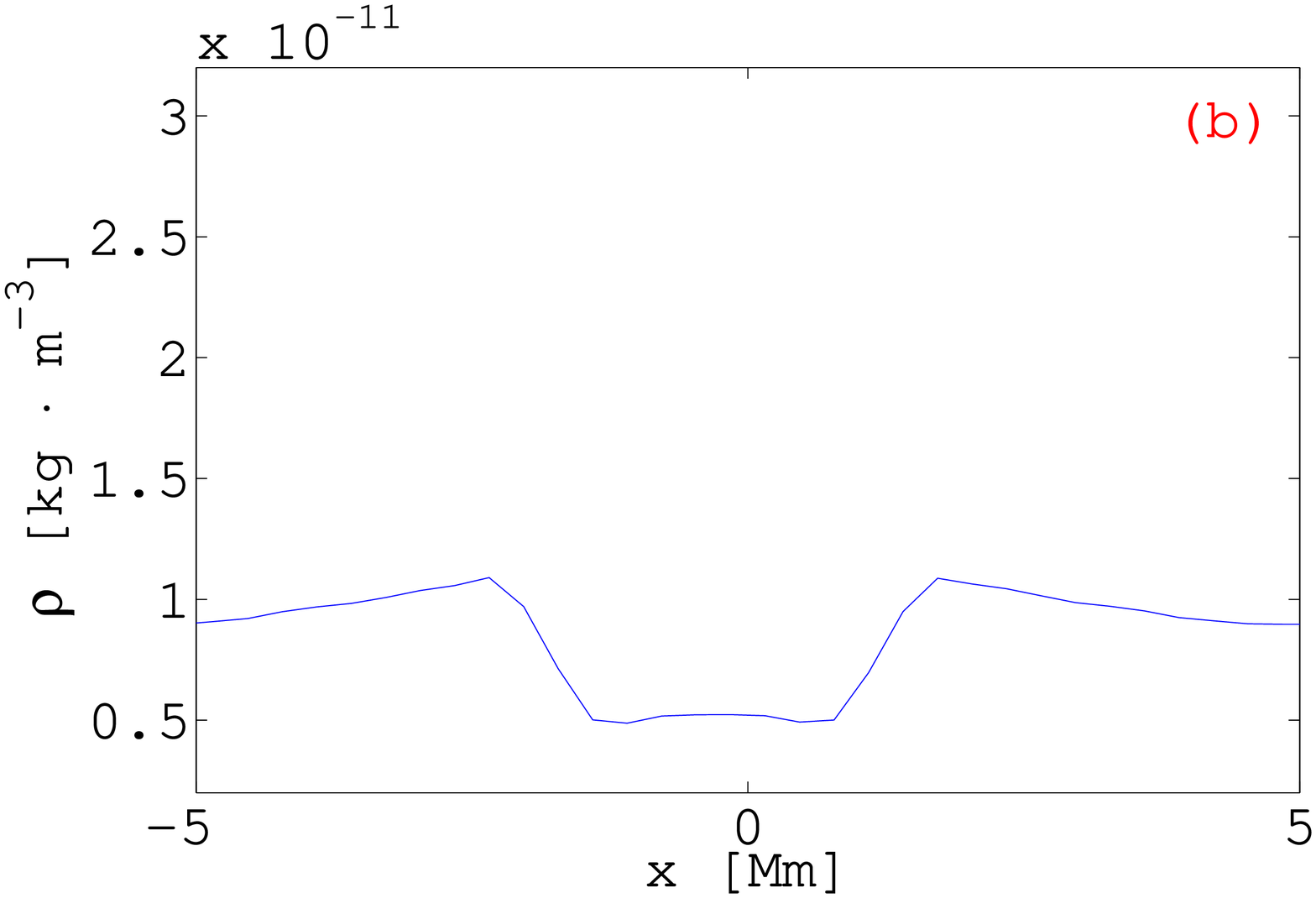}
}
\vspace*{-0.5cm}
\end{center}
\begin{center}
\footnotesize{$t=175~\mathrm{s}$}
\end{center}
\begin{center}
\vspace*{-0.3cm}
\hspace*{-0.5cm}
\mbox{
\includegraphics[scale = 0.27]{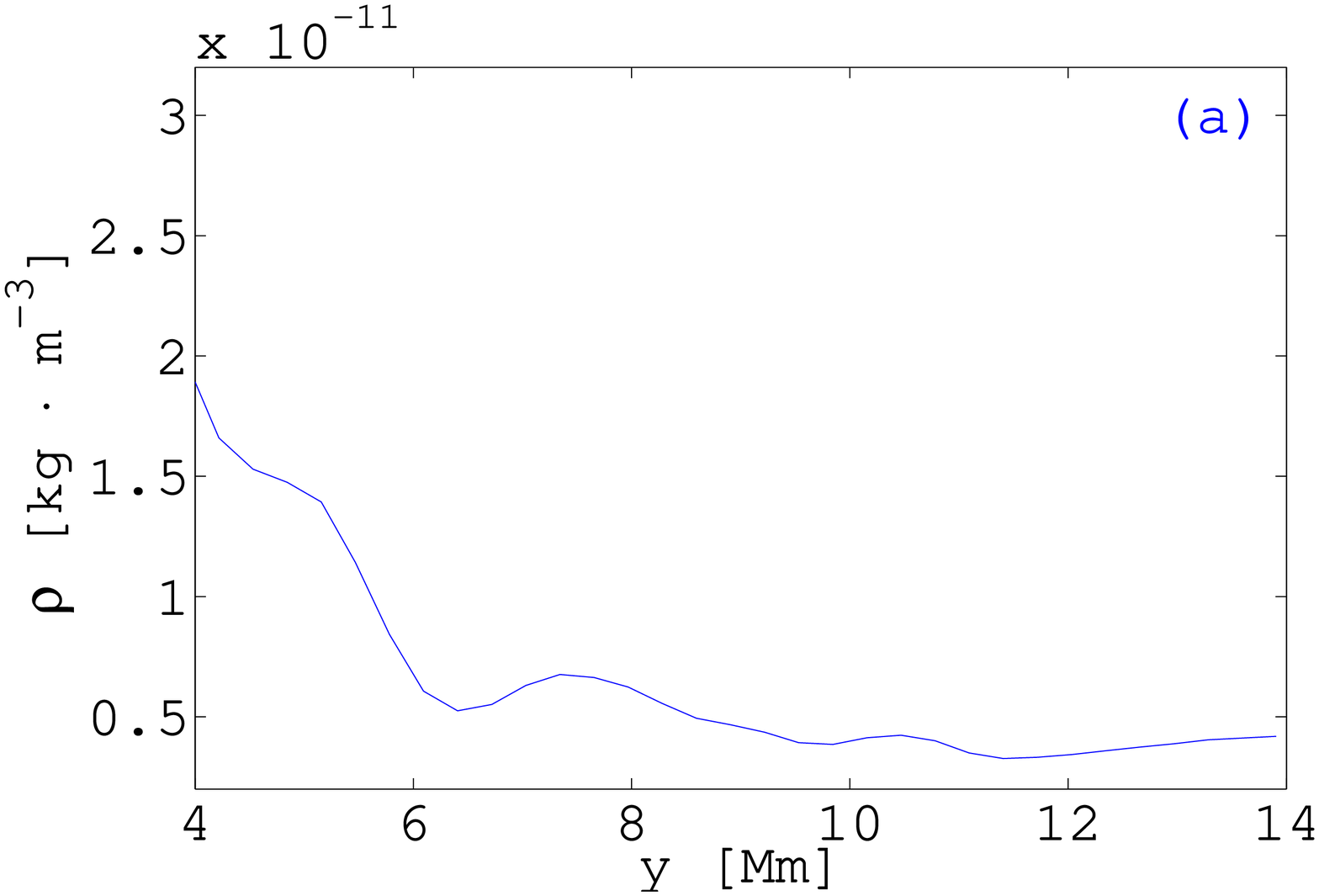}
\includegraphics[scale = 0.27]{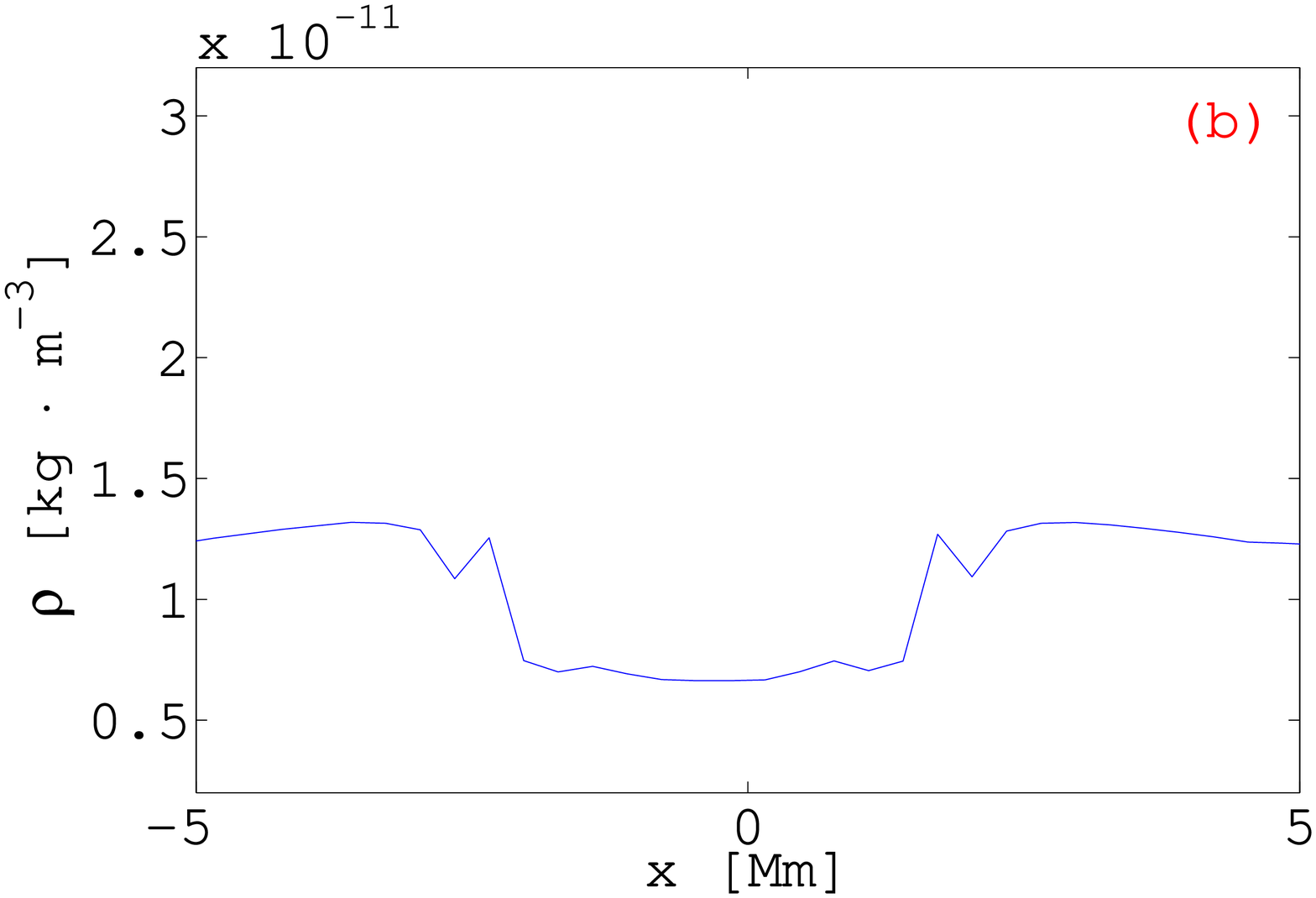}
}
\vspace*{-0.5cm}
\end{center}
\caption{Density profiles along the vertical blue (a) and horizontal red (b) lines in Fig.~\ref{Fig8}.} \label{Fig9}
\end{figure*}

In Fig. \ref{Fig8} we show in detail the densities during the coalescence process, again together with the vertical (blue dashed line) and horizontal (red dashed line) slices. In Fig. \ref{Fig9} we present the slices, which are, similarly as in Fig. \ref{Fig4}, in the positions of the axis of symmetry $(x=0.0~\mathrm{Mm})$ and the position where we find the maximum of mass density on this axis, i.e. $y\approx 7-9~\mathrm{Mm}$. In panel (a) for time $125~\mathrm{s}$ the hump in mass density represents the plasmoid moving downwards to the magnetic arcade, see the movie. At time $135~\mathrm{s}$ we observe the newly formed arcade which starts to oscillate, see the moving humps in panels (a) for times $155$ and $175~\mathrm{s}$ (compare with the temperature in Fig. \ref{Fig13}). Contrary to panels (b) displayed in Fig. \ref{Fig4}, the mass density in the center of the current sheet is lower than in the surrounding plasma. This is caused by the fact that the plasmoid is more dense than the plasma in the place where it was formed. But, because of the gravitational, downward acceleration of the plasma jet, the plasmoid moves down to lower altitudes of the solar atmosphere (transition region), where the density is higher, thus reversing the density contrast.

\begin{figure}[h!]
\hspace{-0.5cm}
 \includegraphics[scale = 0.23]{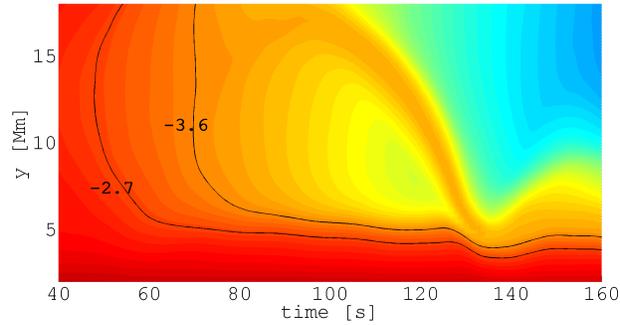}
 \caption{The temporal map of the vector potential at heights $2-18~\mathrm{Mm}$ above the solar surface. Here we present the evolution of the arcade magnetic field lines having the vector potential $A=-2.7$ and $A=-3.6$, showing its decrease in height in the early phases.}
 \label{Fig10}
\end{figure}

\begin{figure}[h!]
\hspace{-0.5cm}
 \includegraphics[scale = 0.23]{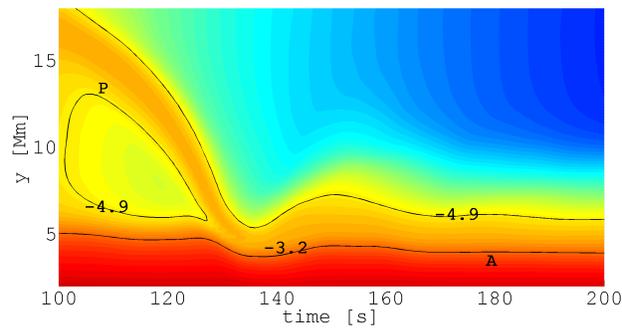}
 \caption{The temporal map of the vector potential at heights $2-18~\mathrm{Mm}$ above the solar surface. Here we present the evolution of the height of the magnetic field lines at the top of the arcade, corresponding to the vector potential $A=-3.2$ and $A=-4.9$, showing the oscillation of the arcade (A) after its merging with the plasmoid (P).}
 \label{Fig11}
\end{figure}

In Figs. \ref{Fig10} and \ref{Fig11} we present again the temporal maps of the vector potential at heights $2-18~\mathrm{Mm}$ above the solar surface. Similar to Fig. \ref{Fig5}, we draw selected vector potentials (magnetic field lines), showing clearly the oscillations of the top of the arcade.

Figure~\ref{Fig10} shows a descent of the arcade top (determined by the magnetic field line in the current sheet axis with the vector potential $A= -2.7$ and $A= -3.5$) at the early stage of the magnetic field reconnection $(50-80~\mathrm{s})$ in the vertical and gravitationally stratified current sheet. This process was already proposed to explain the downward motion of the X-ray source at the beginning of the 3 November 2003 flare \citep{2006A&A...446..675V}.

Figure~\ref{Fig11} shows the interaction of the plasmoid with the arcade and
clearly shows the subsequent arcade oscillations. The plasmoid (P) and arcade
(A) are determined by two magnetic field lines at the current sheet axis with
the vector potential $A = -3.2$ and $A = -4.9$. As seen here, after the
interaction of the plasmoid with the arcade, the arcade starts to oscillate
with a $35~\mathrm{s}$ period. Moreover, it can be seen that oscillations of
different layers of the arcade are not fully synchronized, as is evidenced by
the phase lag between the maxima of the arcade oscillations expressed by the
lines with $A = -3.2$ and $A = -4.9$.

\begin{figure*}[h!]
\begin{center}
\hspace*{-0.5cm}
\mbox{
\includegraphics[scale = 0.21]{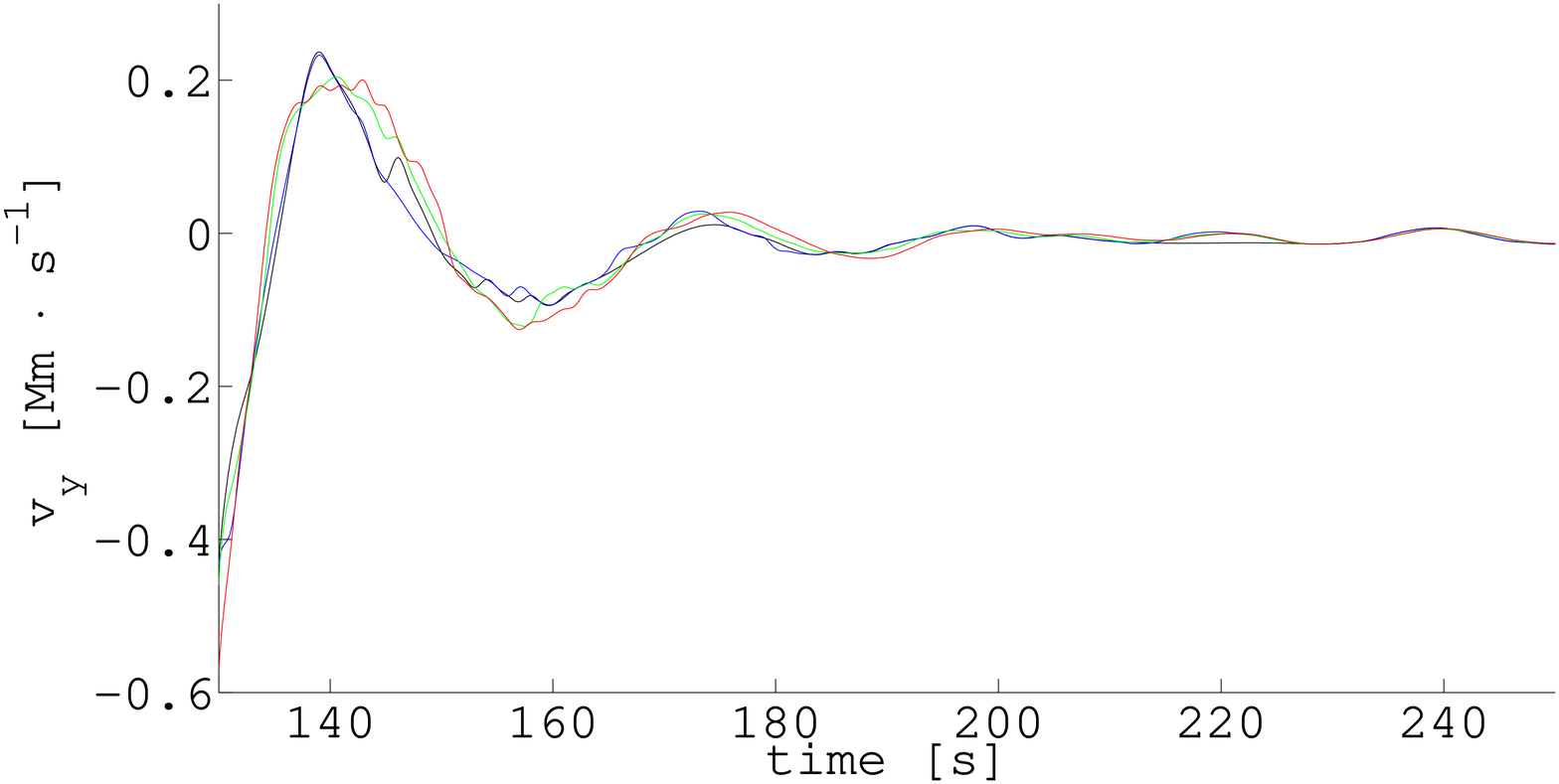}
\hspace{0.cm}
\includegraphics[scale = 0.21]{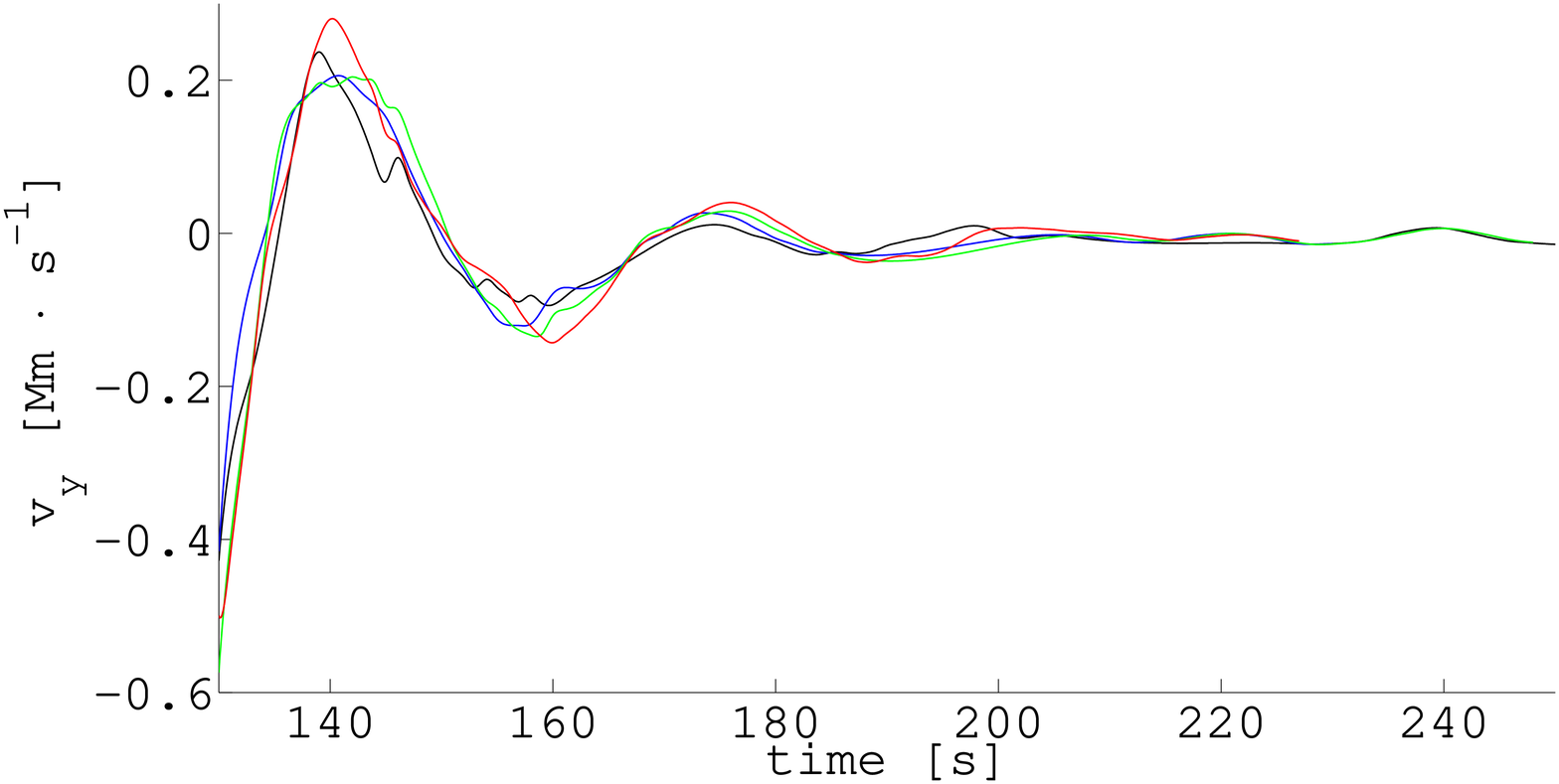}
} \caption{Temporal evolution of the $y$ component of the velocity -- $v_y$ at the selected magnetic field line with the vector-potential $A=-3.8$, showing the standing waves created after the interaction of the plasmoid with the magnetic arcade. In the left part of the figure the black, blue, green and red lines correspond to the detection positions $x = 0.0; -0.3125; -0.625$ and $-0.9375~\mathrm{Mm}$. On the other hand, in the right part the black, blue, green and red lines correspond to the positions $x = 0.0; +0.3125; +0.625$ and $+0.9375~\mathrm{Mm}$. The $y$ coordinates are placed along the selected magnetic field line.}
\label{Fig12}
\end{center}
\end{figure*}

\begin{figure*}[h!]
 \hspace{-0.5cm}
 \begin{center}
 \includegraphics[scale = 0.22]{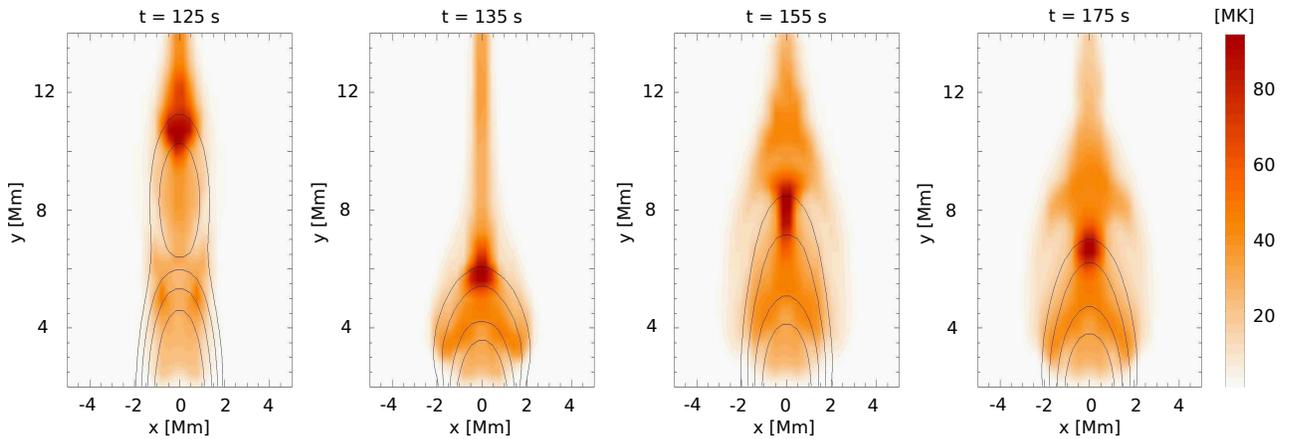}
 \caption{The time evolution (for $t=125, 135, 155, 175~\mathrm{s}$) of the temperature in the region of the coalescence of the plasmoid with the magnetic arcade.}
 \label{Fig13}
 \end{center}
\end{figure*}

In Fig. \ref{Fig12}, we present the temporal evolution of the $y$ component of
the velocity -- $v_y$ in seven detection points along the selected magnetic
field line with the vector-potential $A=-3.9$. On the left panel of Fig.
\ref{Fig12}, we show the signals from the left side of the arcade and on the
right panel signals from the right side of the arcade. The colours correspond
to the positions (depending on the side of symmetry -- left or right) as
follows: black $(0.0~\mathrm{Mm})$, blue $(-0.3125~\mathrm{Mm};
0.3125~\mathrm{Mm})$, green $(-0.625~\mathrm{Mm}; 0.625~\mathrm{Mm})$ and red
$(-0.9375~\mathrm{Mm}; 0.9375~\mathrm{Mm})$, respectively. As can be seen here,
the oscillations are synchronized at both sides of the arcade and quickly
damped. Here, similarly as in the case of the damping of the
oscillating plasmoid, we think that this damping is caused by a strong coupling
between the oscillating flare arcade and plasma outflows from the above located
magnetic reconnection. We estimate the period of the wave signal as
$\tau_\mathrm{arc}\approx 35~\mathrm{s}$. Here as well, the estimated wave
period corresponds to the formula $\tau_\mathrm{arc} = L / c_\mathrm{A} =
38.9~\mathrm{s}$, where $L$ is a length of the loop and $c_\mathrm{A}$ is
average Alfv\'en speed in the loop. A very important finding from our numerical
simulation of the collision of a plasmoid and a magnetic arcade is that the
impinging plasmoid can excite a transverse wave in the arcade as seen from both
parts of Fig. \ref{Fig12}. We can infer that standing waves are observed in our
simulations because the peaks of the wave signal are synchronized very well
between symmetric locations, as seen in Fig. \ref{Fig12}. These standing,
transverse waves may be responsible for the generation of the observed QPPs in
solar flares.

To complete this part of our study, we show the time evolution of the
temperature for different times $125,135,155$ and $175~\mathrm{s}$ in Fig.
\ref{Fig13}. We shifted the bottom boundary to higher altitudes of the solar
atmosphere $(4~\mathrm{Mm})$ because of better readability in a linear
temperature scale. We can clearly see that the plasmoid coalescence causes a
temperature enhancement up to $\approx 80~\mathrm{MK}$. Note, that in real
flares, especially in high-temperature regions there are deviations from our
simplifying assumption ($R_{\rm{loss}} + T_{\rm{cond}} + H = 0$), and thus we
expect that the real temperature will be lower. This temperature blob moves
slowly down and after the collision and merging of the plasmoid with the arcade
this blob starts to oscillate following the position of the top of the newly
created magnetic arcade. During the observed oscillations, there is practically
no change of the maximum of the temperature, so it remains almost at the same
value.

Using relations according to \cite{2004psci.book.....A} we estimated
the radiative cooling times $\tau_\mathrm{R}$ (in optically thin approximation)
and conductive cooling times $\tau_\mathrm{C}$ (parallel to the magnetic field
lines) in several selected positions at the time 175 s (the time with the
maximal temperature in the arcade top source). We found that at this instant
the radiative cooling time for the plasmoid is $\tau_\mathrm{R} \approx
19~\mathrm{s}$, for the arcade top source is $\tau_\mathrm{R} \approx
15~\mathrm{s}$ and for the upper chromosphere $\tau_\mathrm{R} \approx
0.3~\mathrm{s}$. On the other hand, the conductive cooling times in the
parallel direction to the magnetic field lines for the arcade top source is
$\tau_\mathrm{C} \approx 0.025~\mathrm{s}$ and for the transition region
$\tau_\mathrm{C} \approx 12~\mathrm{s}$. The conductive cooling times in the
perpendicular direction to the magnetic field lines are typically several
orders longer. As can be seen, the shortest times are the conductive cooling
time at the hot arcade top (extreme in the present computation) and the
radiative cooling time at the upper chromosphere. It looks that the temperature
inside the arcade will be in very short time smoothed and the upper
chromosphere will rapidly cool to a very low temperature. But in reality, there
is an additional heating, which is not well known and keeps for example the
chromosphere and the transition region stable for long time. For this reason we
used the assumption that the radiative and conductive losses are compensated by
the assumed additional heating. This is a weak aspect of our simulations,
especially for the hot arcade top. But, to add the terms for radiative and
conductive losses to MHD equations without the additional heating term does not
solve the problem according to our opinion. We found a hot source in the flare
arcade top, as observed, but its temperature is much higher than that in
observed loop-top sources. In this case, we think that our assumption about
compensation of radiative and conductive losses is not valid.

The high temperature near the loop top may result in associated loop-top
sources of flares. Therefore, the current mechanism of a merging plasmoid with
a magnetic arcade could compete with, operate simultaneously, or help the
previously proposed mechanism of \citet{2016ApJ...833...36F}. They proposed that the
inverse Compton scattering together with Kelvin-Helmholtz turbulence could lead
to long-lived loop-top sources.

As for the energetic aspects of the merging, the kinetic, potential and
magnetic energy of the plasmoid is completely absorbed in the arcade, because
there is no reflected plasmoid in Fig.~\ref{Fig12} (bottom right panel). Thus,
the plasmoids may form the missing link for energy transport in flares. The
plasmoids could be the agent transporting a large part of the energy from the
flare site to the arcade. From there, the waves in the arcade may take over the
energy transport. After coalescence with the underlying magnetic arcade, the
plasmoid energy is now stored and transported in the arcade in the form of fast
or Alfv\'en waves, following the scenario outlined by
\citet{2008ApJ...675.1645F}. This energy is then dissipated in the lower
atmospheric layers by non-ideal effects \citep[see,
e.g.][]{2016ApJ...818L..20R}. Such a scenario is a hybrid between previously
proposed energy transport mechanisms in flares.

\section{Conclusions}
We numerically studied the oscillatory and wave processes in a gravitationally
stratified current sheet, together with a realistic solar atmosphere structured
as the VAL-C model, using the 2D time-dependent non-ideal (resistive) MHD
equations solved by the FLASH numerical code, which implements AMR.

We can summarize our results as follows. We generated an initial gaussian pulse
with the horizontal component of the velocity in the current sheet, which
triggers the primary reconnection. Subsequently it leads to the formation of a
number of plasmoids. These plasmoids, under the gravitational and buoyancy
forces start to move upwards or downwards. After some time, the formed
plasmoids collide with each other as well as with the magnetic arcade formed in
the lower altitudes of the solar atmosphere. We found that the magnetic arcade
shrinks in the very early phases of evolution, in agreement with observations
of downward motion of the X ray source at the beginning of the 3 November 2003
flare \citep{2006A&A...446..675V}.

As a consequence of the plasmoid collisions, they form very easily much bigger
plasmoids which oscillate. We observed in our simulation such a plasmoid with
an  oscillation period of $\approx 25~\mathrm{s}$. This period is compatible
with the calculated Alfv\'en travel time within the plasmoid. In line with the
simulations made
by~\cite{2000A&A...360..715K,2007A&A...464..735K,2010A&A...514A..28K}, we
expect that this oscillation process can generate the drifting pulsating
structure (DPS) with a "wavy" appearance on the radio spectrum
\citep{2016CEAB...40...93K}.

Moreover, for the first time, we have shown that plasmoids impacting a magnetic
arcade can efficiently generate transverse waves. The vertically polarised
transverse waves have a period of $35~\mathrm{s}$, which is once again
compatible with Alfv\'en travel times in the arcade. The excitation of these
transverse waves could be very important for explaining QPPs and energy
propagation in a solar flare. We propose that plasmoids may carry away part of
the energy from the reconnection site, which is consequently efficiently
deposited in the underlying arcade, leading to wave dissipation in the
(non-ideal) lower layers.

In our simulations we use the simplified assumption that the radiative and
conductive losses are compensated by the heating. This assumption is correct in
the initial equilibrium flare atmosphere. In the perturbed atmosphere,
deviations from this assumption certainly exist, especially in the very
hot source at the arcade top. However, oscillations of the plasmoid and
arcade found in the present computations are very similar to those observed.
For example, see Figure 5 and 6 in the paper
by~\citet{2010SoPh..266...71K}, where the plasmoid and its oscillation in the
brightness temperature and area are shown. The period of this oscillation was
about $40~\mathrm{s}$. On the other hand, in Figure 6 (panel c) in the paper
by~\citet{2006A&A...446..675V} it can be seen that the velocity of the X-ray
loop-top source shows a damped oscillation at the very beginning of the flare.
Thus, these observations can be considered as supporting our results.

The oscillations of the arcade and plasmoid were strongly damped. Analyzing
this process we found that the oscillating plasmoid as well as the oscillating
arcade are not isolated oscillating systems. Both are strongly coupled with
processes in the vertical current sheet through plasma flows. The downward
oriented plasma flow, carrying plasmoids from the reconnection towards the arcade,
not only triggers the arcade oscillations, but the flow
damps this arcade oscillation subsequently. Similarly, the plasma flows in the vicinity of
the oscillating plasmoid damp its oscillation.

Finally, we found that in both processes (merging of two plasmoids and
interaction of a plasmoid with a flare arcade) the internal structure of the
plasmoids as well as the arcades is very complex. Even an oscillating region
with a very hot plasma close to the arcade top was generated. However, for a
more realistic description of these hot plasma regions the radiative and
conductive losses need to be included into simulations, which constitutes
future work. Furthermore, we will additionally add numerical resolution in the
arcade regions, to allow for a better characterisation of the wave nature.

In our numerical simulations, we use a constant resistivity over the entire
simulation region. For this reason, we have plasmoids and a magnetic arcade of
comparable sizes and as a consequence of their collision standing waves are
triggered. It may be that smaller plasmoids impacting on larger arcades would
rather excite propagating, transverse waves. This could also be the subject of
future research.

\acknowledgments The authors thank the referee for constructive comments that improved the paper. P.~J., M.~B. and M.~K. acknowledge support from Grants 16-13277S and 17-16447S of the Grant Agency of the Czech Republic. T.~V.~D. was supported by an Odysseus grant of the FWO Vlaanderen, the IAP P7/08 CHARM (Belspo) and the GOA-2015-014 (KU~Leuven). The results were inspired by discussions at the ISSI Bern and at ISSI Beijing. The authors also express their thanks to Prof. Kris Murawski for valuable discussions.\\
The FLASH code used in this work was developed by the DOE-supported ASC/Alliances Center for Astrophysical Thermonuclear Flashes at the University of Chicago.




\newpage

\end{document}